\documentclass{ametsocV6.1}

\usepackage{xurl}
\usepackage{cancel}
\usepackage{natbib}
\usepackage{placeins}

\definecolor{revAcolor}{RGB}{0,114,178}
\newif\ifhighlightrevisions
\highlightrevisionsfalse
\ifhighlightrevisions
  \DeclareRobustCommand{\revA}[1]{{\color{revAcolor}#1}}
\else
  \DeclareRobustCommand{\revA}[1]{#1}
\fi

\defcitealias{Kim+An2021JCLI}{KA21}
\defcitealias{Vialard+2025RevGeophys}{V25}
\defcitealias{Han+2026JCLI}{H26}

\nolinenumbers
\let\internallinenumbers\relax

\title{A Stochastic-Thermodynamic Constraint on the Seasonal Phase Locking of the El Ni\~no--Southern Oscillation}

\authors{
    Yuki Yasuda\aff{a}\correspondingauthor{Yuki Yasuda, yuki.yasuda@jamstec.go.jp}
    and Tsubasa Kohyama\aff{b}
}

\affiliation{
    \aff{a}{Research Institute for Earth and Information Sciences, Japan Agency for Marine-Earth Science and Technology}\\
    \aff{b}{Department of Information Sciences, Ochanomizu University}
}

\abstract{
    We investigate the seasonal phase locking of the El Ni\~no--Southern Oscillation (ENSO) in a linear stochastic recharge oscillator (SRO), a damped oscillator with additive noise and a time-dependent growth rate.
    Phase locking is reflected in the seasonality of the sea surface temperature anomaly (SSTA) \revA{variance.}
    In general, energy drives such a change, whereas entropy \revA{constrains} whether it \revA{can occur.}
    \revA{An energy-based perspective on phase locking can thus be complemented by an entropy-based constraint.}
    We quantify this constraint using a thermodynamic uncertainty relation (TUR), a fundamental inequality in stochastic thermodynamics.
    The TUR \revA{gives an upper bound on} the tendency of the SSTA variance \revA{in terms of} the partial entropy production rate, which is dominated by the ratio of forward and backward transition probabilities and quantifies the irreversibility of SSTA transitions.
    \revA{The seasonal variation of this irreversibility is governed by the growth rate, which reaches its extrema in boreal autumn and late winter.}
    \revA{At both times, the entropy production rate becomes large, relaxing the TUR constraint so that the SSTA variance can peak in boreal winter, consistent with observed ENSO phase locking.}
    \revA{We further show that the entropy production rate in the SRO splits into this seasonal component and the irreversibility associated with the recharge--discharge cycle of ENSO.
    Based on these results, we discuss a hypothesis for real ENSO.}
    If irreversibility were interpreted as dissipated energy, the \revA{TUR} constraint on ENSO growth and decay would require this dissipation to be exported from the equatorial Pacific.
}

\begin{document}

\maketitle

\section{Introduction}\label{sec:introduction}

The El Ni\~no--Southern Oscillation (ENSO) is the dominant mode of interannual variability in the tropical Pacific and influences the global atmosphere and ocean \citep[e.g.,][]{Bjerknes1969MWR,Timmermann+2018Nature,Taschetto+2020inbook,Trenberth20inbook}.
ENSO is characterized by sea surface temperature anomalies (SSTAs) in the eastern equatorial Pacific, which tend to peak in boreal winter during El Ni\~no and La Ni\~na (the warm and cold phases of ENSO, respectively).
This seasonal peak, known as phase locking, remains a key challenge for climate models to reproduce \citep[e.g.,][]{Chen+Jin2021JCLI,Liao+2021DSR,Yang+2023NPJ} and motivates a deeper understanding of the underlying dynamics.

The recharge oscillator \citep[RO;][]{Jin1997aJAS} is an idealized model widely used to study the temporal evolution of ENSO, including phase locking; for recent reviews, see \citet{Jin+2020inbook} and \citet{Vialard+2025RevGeophys}.
With its parameters estimated from observational or climate model data, the RO is inferred to be a damped oscillator driven by stochastic noise \citep{Burgers+2005GRL,Stein+2010JCLI,Stein+2014JCLI,Chen+Jin2020JCLI,Chen+Jin2022GRL,Kim+An2021JCLI,Han+2026JCLI}, although robustly distinguishing a damped from a self-sustained regime may be difficult \citep{Weeks+Tziperman2025GRL}.
In this realistic parameter regime, the seasonality of the SSTA growth rate is considered the dominant cause of phase locking \citep{Stein+2010JCLI,Stein+2014JCLI,Chen+Jin2020JCLI,Chen+Jin2021JCLI,Chen+Jin2022GRL,Kim+An2021JCLI,Li+2025GRL,Han+2026JCLI}.
Other factors include the seasonality of the oscillation frequency \citep{Stein+2010JCLI,An+Jin2010TellusA,Chen+Jin2022GRL}, external forcing by the annual cycle \citep{Stein+2014JCLI}, and nonlinearity \citep{Chen+Jin2020JCLI,Kim+An2021JCLI}, but none has been well established as the dominant factor.
The former two are omitted in most recent RO formulations, whereas the remaining factor, nonlinearity, is still necessary to reproduce the asymmetry between El Ni\~no and La Ni\~na \citep{Chen+Jin2020JCLI,Kim+An2021JCLI,Han+2026JCLI}.

\citet{Kim+An2021JCLI} (hereafter \citetalias{Kim+An2021JCLI}) examined the noise-driven RO and proposed the seasonal energy index (SEI) as a diagnostic measure of phase locking.
In the noise-driven RO (stochastic RO; SRO), phase locking is reflected in the seasonality of SSTA variance: ENSO events peak in boreal winter, when the variance is largest.
To analyze this seasonality, \citetalias{Kim+An2021JCLI} noted that the SRO has two degrees of freedom and interpreted SSTA as momentum in classical mechanics.
In this analogy, the SSTA variance corresponds to kinetic energy.
They showed that the time-integrated growth rate determines the seasonal modulation of SSTA variance and defined the SEI based on this integral.
However, the SEI is mainly discussed from a deterministic viewpoint, and a stochastic analysis of phase locking remains less developed in the SRO.

Such a stochastic analysis is naturally framed in terms of entropy, whose \emph{production} quantifies the irreversibility of a stochastic process \citep[e.g.,][]{Seifert2005PRL}.
\revA{In general, entropy constrains whether a change in a physical system can occur, whereas energy is required to drive the change.}
\revA{This fact suggests that the energy-based perspective on phase locking can be complemented by an entropy-based constraint.}
Quantifying such a constraint requires a framework that connects entropy production to fluctuations.

Over the past two decades, the study of fluctuations in nonequilibrium systems has advanced within the framework of \emph{stochastic thermodynamics} \citep[e.g.,][]{Seifert12RepProgPhys,Parrondo+15NatPhys,Peliti+Pigolotti21Book,Shiraishi23Book}.
Despite its name, this framework also applies to nonthermal systems and can be used for multivariable nonlinear systems driven by stochastic noise.
Such noise is present in climate dynamics, where small-scale motions in the atmosphere and ocean can be modeled as stochastic forcing \citep[e.g.,][]{Hasselmann1976Tellus,Franzke+2015WIREClimChange}, and thus stochastic thermodynamics may be applicable.
Indeed, \citet{Yasuda+Kohyama2025JCLI} applied this framework to interannual-to-decadal synchronization between the Gulf Stream and the Kuroshio \citep{Kohyama+21Science} and suggested that information-to-energy conversion \citep{Toyabe+2010NatPhys,Koski+14PNAS} may occur even in the climate system.
Such new physical insights may emerge through stochastic thermodynamics, but this framework is relatively new and has \revA{not been sufficiently applied to the climate system.}

\revA{Previous studies have analyzed irreversibility in tropical sea surface temperature (SST) variability \citep{Weiss07,Weiss09}, ENSO \citep{Weiss+20,Queiros24}, the Madden--Julian oscillation \citep[MJO;][]{Weiss+20,Benincasa26}, and weather patterns over western Europe \citep{Vannitsem23}.}
\revA{Most of these studies modeled the fluctuations with time-independent coefficients \citep{Weiss07,Weiss09,Weiss+20,Queiros24,Benincasa26}, which cannot represent seasonal phase locking.}
\revA{In addition, none of them used a thermodynamic uncertainty relation (TUR).}
\revA{TURs are recently proposed fundamental inequalities in stochastic thermodynamics that bound fluctuations by entropy production (a measure of irreversibility) and can hold at each time even for systems with time-dependent coefficients \citep{Barato+Seifert2015PRL,Horowitz+Gingrich2020NatPhys,Tanogami+2023PRR}.}

We analyze the SRO using \revA{a TUR} and show that ENSO phase locking is \revA{subject to a stochastic-thermodynamic constraint.}
\revA{In the SRO, sufficiently large irreversibility, quantified by the partial entropy production rate, is necessary for the seasonal change in SSTA variance that reflects phase locking.}
Section~\ref{sec:sro} introduces a linear SRO and its parameter values.
Section~\ref{sec:tur} reviews the TUR \revA{and discusses its limitations.}
Section~\ref{sec:results} applies the TUR to the SRO, and Section~\ref{sec:discussion} then discusses the implications for real ENSO.
Finally, Section~\ref{sec:conclusions} presents the conclusions.

\section{Linear stochastic recharge oscillator (SRO)} \label{sec:sro}

\begin{figure}[t]
    \centering
    \noindent\includegraphics[width=8cm]{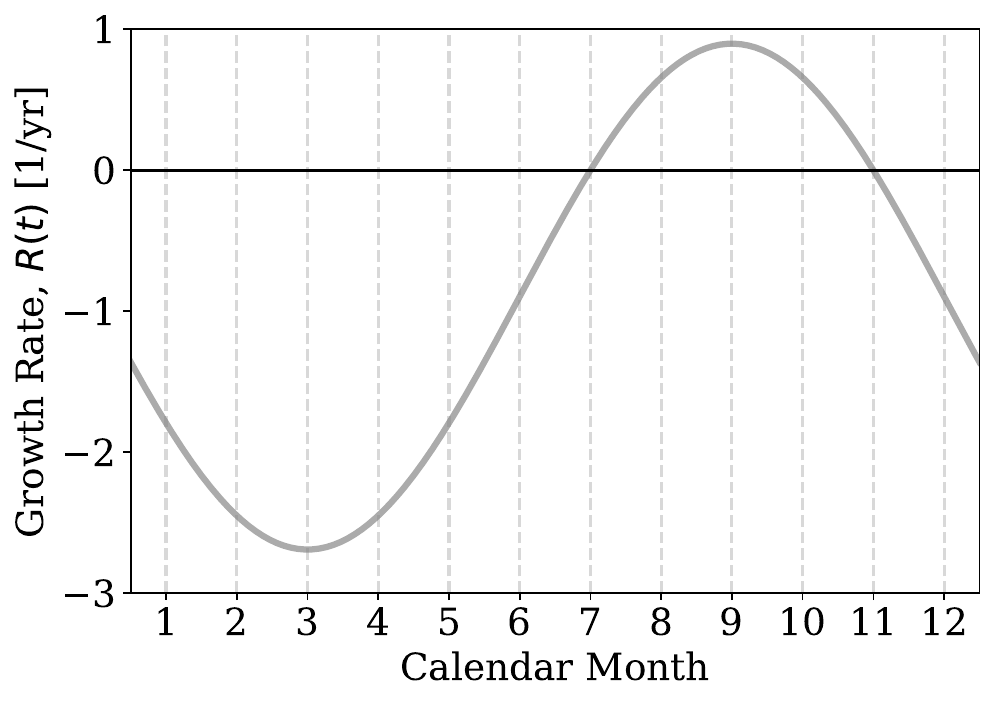}
    \caption{
        Annual cycle of the SSTA growth rate $R(t) = R_0 - R_a \sin(\omega t + \phi)$, with parameter values from \citetalias{Kim+An2021JCLI} (Table~\ref{table:sro-parameters}).
        The frequency $\omega = 2\pi$ and phase $\phi = 2\pi/24$ are chosen so that $R(t)$ peaks in September.
    }
    \label{fig:annual-cycle-growth-rate-KA21}
\end{figure}

We use the following linear SRO \citep[e.g.,][]{Vialard+2025RevGeophys}:
\begin{align}
  \frac{{\rm d} T(t)}{{\rm d} t}&=R(t) T(t)+F_1 h(t)+\sigma^T \xi^T(t), \label{eq:linear-sro1}\\
  \frac{{\rm d} h(t)}{{\rm d} t}&=-\varepsilon h(t)-F_2 T(t)+\sigma^h \xi^h(t), \label{eq:linear-sro2}\\
  R(t)&=R_0 - R_a \sin(\omega t + \phi), \label{eq:linear-sro3}\\
  \left\langle\xi^T(t)\xi^T(t')\right\rangle &= \left\langle\xi^h(t)\xi^h(t')\right\rangle = \delta(t-t'), \label{eq:linear-sro4}\\
  \left\langle\xi^T(t)\xi^h(t')\right\rangle &= 0,\label{eq:linear-sro5}
\end{align}
where $T$ denotes the SSTA in the eastern equatorial Pacific, and $h$ denotes the thermocline depth anomaly over the entire equatorial Pacific, corresponding to ocean heat content.
The Gaussian white noises $\xi^T$ and $\xi^h$ are mutually independent, and the angle brackets $\langle \cdot \rangle$ denote the ensemble average over noise realizations.
All coefficients are constant except $R(t)$.
The SSTA growth rate $R(t)$ encapsulates multiple physical processes, including the Bjerknes feedback \citep[e.g.,][]{Jin+2020inbook}.
It has an annual cycle that peaks around September (Fig.~\ref{fig:annual-cycle-growth-rate-KA21}).
The parameter $R_0$ is the annual mean, and $R_a$ is the amplitude of the annual-cycle component.
We discuss causes of the annual cycle in $R(t)$ later (Section~\ref{sec:discussion}\ref{subsec:interpretation}).
Our notation follows a recent review \citep{Vialard+2025RevGeophys}.
We may also write the time $t$ as a subscript when there is no confusion, following the convention for stochastic processes [e.g., $R_t = R(t)$].

The parameters of the linear SRO are taken from \citetalias{Kim+An2021JCLI}, \citet{Vialard+2025RevGeophys}, or \citet{Han+2026JCLI} (Table~\ref{table:sro-parameters}).
The \citetalias{Kim+An2021JCLI} parameter set is our main case, whereas the other two sets are used to test the robustness of our results.
\revA{\citetalias{Kim+An2021JCLI} estimated the parameters by least-squares fitting to $T$ from ERSSTv5 and $h$ from the SODAv2.2.4 (1958--2010) and GODAS (2011--2016) reanalyses.
\citet{Vialard+2025RevGeophys} fitted the parameters to $T$ and $h$ from the ORAS5 reanalysis for 1958--2020.
\citet{Han+2026JCLI} first estimated the parameters by regression on $T$ from HadISST and $h$ from the ORAS5 reanalysis for 1970--2020 and then refined them through parameter-sweep experiments to reproduce observed ENSO statistics, such as the seasonal standard deviation.}
Here, we use only the parameter values that appear in the linear SRO [Eqs.~(\ref{eq:linear-sro1})--(\ref{eq:linear-sro5})], not those of any additional terms \citep[e.g., nonlinear terms in][]{Han+2026JCLI}. 

\begin{table}[t]
    \caption{
        Parameter values of the linear SRO [Eqs.~(\ref{eq:linear-sro1})--(\ref{eq:linear-sro5})].
        The columns KA21, V25, and H26 correspond to the parameter sets of \citet{Kim+An2021JCLI}, \citet{Vialard+2025RevGeophys}, and \citet{Han+2026JCLI}, respectively.
        The entries for $T$ and $h$ indicate the averaging regions for SSTA and thermocline depth, respectively.
        The V25 value of $R_a$ is set to $2.0 \lvert R_0 \rvert$ for reference, as in the KA21 experiments.
    }
    \label{table:sro-parameters}
    \begin{center}
        \begin{tabular}{llll}
        \hline\hline
            ~ & KA21 & V25 & H26  \\
        \hline
            $T$ region & central--eastern & eastern & eastern \\
            $h$ region & basinwide & basinwide & western \\
            $R_0$ [1/yr] & $-0.898$ & $-1.13$ & 0.360 \\
            $R_a$ [1/yr] & 1.80 & 2.26 & 1.92 \\
            $\omega$ [1/yr] & 6.28 & 6.28 & 6.28 \\
            $\phi$ [rad] & 0.262 & 0.262 & 0.262 \\
            $F_1$ [K/m/yr] & 0.242 & 0.190 & 0.180 \\
            $\varepsilon$ [1/yr] & 0.379 & 0.380 & 1.56 \\
            $F_2$ [m/K/yr] & 11.0 & 17.6 & 17.4 \\
            $\sigma^T$ [K\,yr$^{-1/2}$] & 0.797 & 2.22 & 0.610 \\
            $\sigma^h$ [m\,yr$^{-1/2}$] & 5.82 & 14.4 & 5.54 \\
            BWJ Index (Re Part) [1/yr] & $-0.638$ & $-0.755$ & $-0.600$ \\
            BWJ Index (Im Part) [1/yr] & 1.61 & 1.79 & 1.49 \\
            Intrinsic Period [yr] & 3.90 & 3.51 & 4.23 \\
        \hline
        \end{tabular}
    \end{center}
\end{table}
\FloatBarrier

In Table~\ref{table:sro-parameters}, the Bjerknes--Wyrtki--Jin (BWJ) index represents the eigenvalues of the SRO coefficient matrix at $R(t) = R_0$ \citep{Jin+2006GRL,Lu+2018JCLI,Jin+2020inbook}.
The eigenvalues are complex numbers with negative real parts.
Thus, the SRO is a damped oscillator driven by stochastic noise.
Indeed, even with a time-dependent $R(t)$, the system relaxes toward $T=h=0$ when the noise amplitudes are zero ($\sigma^T=\sigma^h=0$).
Although $R(t)$ can become positive and amplify $T$ through the $R(t)T$ term in Eq.~(\ref{eq:linear-sro1}), this effect is not sufficient to drive the SRO.

The three parameter sets differ mainly in the noise amplitudes and in the values of $R_0$ and $\varepsilon$ (Table~\ref{table:sro-parameters}).
The noise amplitudes $\sigma^T$ and $\sigma^h$ of \citet[][V25]{Vialard+2025RevGeophys} are about three times as large as those of \citetalias{Kim+An2021JCLI} and \citet[][H26]{Han+2026JCLI}.
Moreover, \citetalias{Han+2026JCLI} uses a positive $R_0$ and an $\varepsilon$ value that is about four times as large as those of \citetalias{Kim+An2021JCLI} and \citetalias{Vialard+2025RevGeophys}.
This is because \citetalias{Han+2026JCLI} defines $h$ as the thermocline depth anomaly in the western equatorial Pacific, whereas \citetalias{Kim+An2021JCLI} and \citetalias{Vialard+2025RevGeophys} define it over the entire equatorial Pacific; see \citetalias{Han+2026JCLI} for a discussion of this choice.
Despite these differences, all three parameter sets yield similar results in the analyses below, suggesting that our results are not sensitive to the choice of $h$ or to the specific parameter values.
The case of the \citetalias{Han+2026JCLI} parameter set, in which $h$ is taken in the western equatorial Pacific, is discussed in Appendix~D.

In the linear SRO, El Ni\~no and La Ni\~na are symmetric (e.g., \citetalias{Kim+An2021JCLI}).
By linearity, the sign-reversed state $(-T, -h)$ is also a solution.
Thus, the statistical properties of local maxima and minima are identical.
Although nonlinear terms are necessary to model the El Ni\~no--La Ni\~na asymmetry \citetext{\citealp{Chen+Jin2020JCLI}; \citetalias{Kim+An2021JCLI}; \citealp{Han+2026JCLI}}, their effects are not dominant in reproducing phase locking \citetext{\citealp{Chen+Jin2020JCLI}; \citetalias{Kim+An2021JCLI}}.
Hereafter, we focus mainly on El Ni\~no events (that is, local maxima of $T$) and refer to the linear SRO simply as the SRO.

Our numerical setup follows \citetalias{Kim+An2021JCLI}.
The unit of time is one model year of 360 days (30 days per month).
Numerical integration was performed with the Euler--Maruyama method, using a time step of 1 day ($=1/360$~yr).
The initial conditions were set to $T = 0$~K and $h = 10$~m, although the results are not sensitive to this choice.
Similar results were obtained for Gaussian random initial conditions, because the influence of the initial state decays rapidly under stochastic noise.
The integration length and ensemble size differ across experiments and are given in each figure caption.
We used sufficiently long integrations and large ensembles so that the results are insensitive to these choices.
All results can be reproduced with our public code (see Data Availability).

\begin{figure}[t]
    \centering
    \noindent\includegraphics[width=14cm]{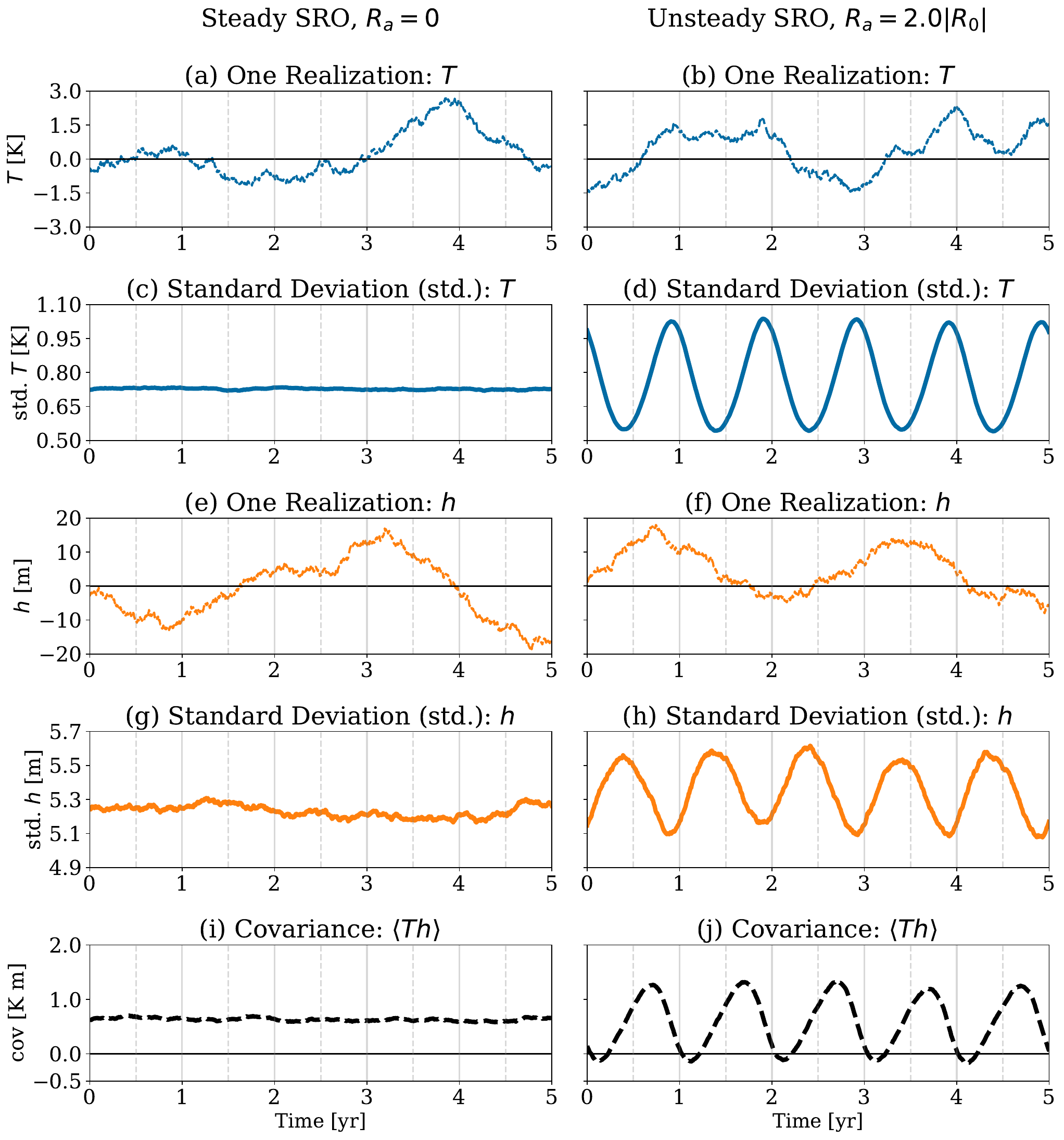}
    \caption{
        Simulation results of the SRO with the \citetalias{Kim+An2021JCLI} parameter set (Table~\ref{table:sro-parameters}).
        \revA{The left column shows the steady SRO with constant growth rate ($R_a = 0$, so $R(t) = R_0$), whereas the right shows the unsteady SRO with seasonally varying growth rate ($R_a = 2.0\lvert R_0 \rvert$).}
        Here, $T$ is the SSTA in the eastern equatorial Pacific, and $h$ is the thermocline depth anomaly over the entire equatorial Pacific.
        Each integration was run for 200 years with 10,000 ensemble members; the last 5 years are shown.
        (a),(b) One realization of $T$.
        (c),(d) Standard deviation of $T$.
        (e),(f) One realization of $h$.
        (g),(h) Standard deviation of $h$.
        (i),(j) Covariance $\langle Th\rangle$.
    }
    \label{fig:sro-time-series-KA21}
\end{figure}

We first confirm that the SRO simulation results are similar to those in \citetalias{Kim+An2021JCLI}.
For comparison, we also numerically integrated a steady SRO with constant growth rate $R(t) = R_0$.
Each integration was run for 200 years with 10,000 ensemble members.
Figure~\ref{fig:sro-time-series-KA21} shows the last 5 years of one ensemble member together with the standard deviation across all members.
For the single member shown, $h$ leads $T$, and $T$ tends to increase after the thermocline has deepened (Fig.~\ref{fig:sro-time-series-KA21}a,b,e,f).
A lagged-correlation analysis using the ensemble confirms that $h$ tends to lead $T$ in both the steady and unsteady SRO (not shown).
The standard deviations vary seasonally: that of $h$ peaks around May and that of $T$ around November (Fig.~\ref{fig:sro-time-series-KA21}d,h).
Between these times, both $h$ and $T$ tend to be positive, which is reflected in the maximum of the covariance $\left\langle T h \right\rangle$ around August (Fig.~\ref{fig:sro-time-series-KA21}j).
These results suggest that the SRO models the phase evolution of ENSO well \citep[e.g.,][]{Vialard+2025RevGeophys}.

A common measure of phase-locking strength is the seasonal variance (or standard deviation) of $T$ \citetext{\citealp{Stein+2010JCLI,Stein+2014JCLI,Chen+Jin2020JCLI}; \citetalias{Kim+An2021JCLI}}.
When the growth rate is constant, the standard deviation of $T$ relaxes to about 0.7~K (Fig.~\ref{fig:sro-time-series-KA21}c), consistent with Fig.~1e of \citetalias{Kim+An2021JCLI}.
As $R_a$ increases, an annual cycle emerges in the standard deviation of $T$.
For $R_a = 2.0 \lvert R_0 \rvert$, the standard deviation of $T$ reaches a maximum of about 1.0~K around November (Fig.~\ref{fig:sro-time-series-KA21}d), consistent with Fig.~1f of \citetalias{Kim+An2021JCLI}.
This seasonal standard deviation corresponds to the concentration of El Ni\~no peaks in a particular season and serves as a phase-locking strength index.
Indeed, although $T$ driven by stochastic noise does not exhibit exact periodicity, its peaks tend to concentrate in November--December (Section~\ref{sec:results}).
In this study, we apply stochastic thermodynamics to the SRO and examine the seasonal variance of $T$.

\FloatBarrier

\section{Thermodynamic uncertainty relation (TUR)} \label{sec:tur}

Stochastic thermodynamics investigates the fundamental properties of fluctuations in nonequilibrium systems \citep{Seifert12RepProgPhys,Parrondo+15NatPhys,Peliti+Pigolotti21Book,Shiraishi23Book}.
Despite its name, it also applies to nonthermal systems.
In fact, the fluctuation theorem, which plays a central role in stochastic thermodynamics, can hold for general nonlinear stochastic systems \citep[e.g.,][]{Harris+Schutz2007JSM}.
As an important consequence of the fluctuation theorem, thermodynamic uncertainty relations (TURs) have been actively studied in recent years \citep{Barato+Seifert2015PRL,Hasegawa+Vu2019PRL,Timpanaro+2019PRL,Horowitz+Gingrich2020NatPhys}.
Applying a TUR to the SRO yields the following inequality for the variance of $T$:
\begin{equation}
    \frac{\dot{\cal P}}{2} \ge \frac{1}{\left(\sigma^T\right)^2\left\langle T^2 \right\rangle} \left( \frac{1}{2} \frac{{\rm d}}{{\rm d}t} \left\langle T^2 \right\rangle \right)^2. \label{eq:tur-shown-for-the-first-time}
\end{equation}
This section introduces this inequality and describes its terms, including the partial entropy production rate $\dot{\cal P}$.

TURs are regarded as fundamental inequalities that \revA{constrain fluctuations by entropy production \citep[e.g.,][]{Horowitz+Gingrich2020NatPhys} and} can hold even for multivariable nonlinear systems or systems driven by multiplicative noise \citep[e.g.,][]{Dechant2019JPA,Tanogami+2023PRR}.
The specific form of the inequality, however, depends on the problem setting; for details, see a review and subsequent studies \citep{Horowitz+Gingrich2020NatPhys,Otsubo+2020PRE,Tanogami+2023PRR}.
Applied to the SRO, the above inequality~(\ref{eq:tur-shown-for-the-first-time}) means that $\dot{\cal P}$ provides an upper bound on the squared tendency of $\left\langle T^2 \right\rangle$, normalized by $\left(\sigma^T\right)^2$ and $\left\langle T^2 \right\rangle$.
The partial entropy production rate $\dot{\cal P}$ reflects the system's irreversibility (Sections~\ref{sec:tur}\ref{subsec:tur-general} and~\ref{sec:results}).
Thus, this inequality links irreversibility to the seasonal variance.

We review a TUR for the subsystem $T$ of the full $(T,h)$ state \citep{Tanogami+2023PRR}.
After introducing the general form of the TUR (Section~\ref{sec:tur}\ref{subsec:tur-general}), we discuss the specific form used in this study (Section~\ref{sec:tur}\ref{subsec:tur-specific}).
\revA{We then apply the TUR to a one-variable version of the SRO and clarify the meaning of irreversibility (Section~\ref{sec:tur}\ref{subsec:tur-one-variable}).}
Finally, we describe limitations of the TUR-based analysis (Section~\ref{sec:tur}\ref{subsec:tur-limitations}).
Both the general and specific forms \revA{of the TUR} are re-derived in Appendix~A.
\citet{Tanogami+2023PRR} obtained a TUR for time-integrated currents and argued that an instantaneous form, without time integration, can be obtained from it.
Alternatively, we use a short-time-limit method \citep{Otsubo+2020PRE} and show that this instantaneous TUR can be derived directly (Appendix~A).

\subsection{General form of the TUR} \label{subsec:tur-general}

\revA{In general, TURs constrain the uncertainty of a \emph{current} (defined below) by giving a lower bound in terms of entropy production \citep[e.g.,][]{Horowitz+Gingrich2020NatPhys}.}
Reducing this lower bound requires increasing entropy production.
Following \citet{Tanogami+2023PRR}, we apply the TUR to the subsystem $T$ (the full system consists of $T$ and $h$).

We first introduce the current ${\cal J}$ associated with $T$:
\begin{align}
    {\cal J} := g(T,h) \circ {\rm d}T,
\end{align}
where $g(T,h)$ is an arbitrary weight function.
In this subsection, we do not restrict the form of $g$, so the TUR is general.
A specific choice, $g = T$, is discussed in Section~\ref{sec:tur}\ref{subsec:tur-specific}.
Stochastic-thermodynamic quantities are denoted by calligraphic symbols (e.g., ${\cal J}$).
The symbol $\circ$ denotes the Stratonovich product, defined as follows \citep[e.g.,][]{Gardiner09Book}:
\begin{equation}
    g(T,h) \circ {\rm d} T := \lim_{{\rm d} t \rightarrow 0} \frac{g(T_{t+{\rm d} t},h_{t+{\rm d} t}) + g(T_{t}, h_{t})}{2} \left(T_{t+{\rm d} t} - T_t\right),
\end{equation}
where subscripts denote time.
With the Stratonovich product, the chain rule holds for the noise-driven, nonsmooth time series $T_t$ \citep[e.g.,][]{Shiraishi23Book}.

The mean and variance of the current characterize the net change in $T$ and the fluctuations of $T$, respectively.
The increment ${\rm d}T$ can be either positive or negative, reflecting the stochastic nature of $T$.
Taking the mean of the current yields
\begin{equation}
  \left \langle {\cal J} \right \rangle = \left \langle g(T,h) \circ {\rm d}T \right \rangle.
\end{equation}
Even when this mean is zero, the increment ${\rm d}T$ itself may still fluctuate.
The variance of the current quantifies the magnitude of these fluctuations:
\begin{align}
  {\rm Var}[{\cal J}] &:= \left \langle {\cal J}^2 \right \rangle - \left \langle {\cal J} \right \rangle^2.
\end{align}

The TUR is the following inequality for the mean and variance of the current:
\begin{align}
    \frac{{\rm Var}[{\cal J}]}{\left \langle {\cal J} \right \rangle^2}  {\rm d} t &\ge \frac{2}{\dot{\cal P}}, \label{eq:tur-general1} \\
    \frac{\dot{\cal P}}{2} &\ge \frac{{\left \langle {\cal J} \right \rangle^2}}{{\rm Var}[{\cal J}]\;{\rm d}t}. \label{eq:tur-general2}
\end{align}
$\dot{\cal P}$ is called the partial entropy production rate and is always nonnegative \citep{Allahverdyan+09JStatMech,Shiraishi+Sagawa2015,Rosinberg+Horowitz2016}; this nonnegativity follows from the second law of information thermodynamics (Appendix~A\ref{app-subsec:second-law}).
It is called \emph{partial} because it represents the entropy production associated only with transitions of $T$.
The inequality~(\ref{eq:tur-general1}) means that, for currents with nonzero mean, the variance normalized by the squared mean of the current (that is, the relative uncertainty) is lower bounded by $2/\dot{\cal P}$.
To reduce this relative uncertainty, one must reduce the lower bound, which requires increasing $\dot{\cal P}$.
The name TUR reflects this interpretation as an inequality about uncertainty \citep{Barato+Seifert2015PRL,Horowitz+Gingrich2020NatPhys}.

In nonequilibrium physics, the TUR is often presented as Eq.~(\ref{eq:tur-general1}), whereas Eq.~(\ref{eq:tur-general2}) is equivalent and can be interpreted as a reformulation in terms of precision.
We use the latter form, Eq.~(\ref{eq:tur-general2}).
The precision, expressed as a signal-to-noise ratio (the squared mean divided by the variance), is upper bounded by $\dot{\cal P}/2$.
In general, stochastic noise induces both the net displacement $\left \langle {\cal J} \right \rangle$ and the ``back-and-forth'' fluctuations ${\rm Var}[{\cal J}]$; their ratio quantifies the relative dominance of the two.
If $\dot{\cal P} = 0$, then the mean current $\left \langle {\cal J} \right \rangle$ is zero and no net displacement occurs (provided that ${\rm Var}[{\cal J}]$ is finite).
In other words, a net displacement must be accompanied by a positive $\dot{\cal P}$.

The partial entropy production rate $\dot{\cal P}$ consists of the following three quantities \citep{Horowitz+Esposito14PhysRevX,Tanogami+2023PRR}:
\begin{align}
    \dot{{\cal P}} &:= \frac{2\dot{\cal Q}}{\left(\sigma^T\right)^2}+\frac{{\rm d} {\cal S}}{{\rm d} t}-\dot{{\cal I}}, \label{eq:def-pdot} \\
    \dot{\cal Q} &:=\left\langle \left( R_t T+F_1 h \right) \circ \frac{{\rm d} T}{{\rm d} t}\right\rangle, \label{eq:def-q} \\
    {\cal S} &:=-\int {\rm d} T_t \; p_t(T_t) \ln p_t(T_t), \label{eq:def-s} \\
    \dot{{\cal I}} &:= \frac{{\cal M}(T_{t+{\rm d}t}:h_t)-{\cal M}(T_t:h_t)}{{\rm d} t}. \label{eq:def-i}
\end{align}
$\dot{\cal Q}$ is an indicator of irreversibility for the change in $T$ and is given by the ratio of the forward and backward transition probabilities \citep{Kurchan1998JPA,Seifert2005PRL,Ito16Book}.
We use the dot notation for \revA{quantities per unit time} that cannot be expressed as total time derivatives (e.g., $\dot{\cal Q}$).
The Shannon entropy ${\cal S}$ represents the uncertainty of the continuous variable $T$ and is a differential entropy defined in terms of the probability density function $p_t(T_t)$ at time $t$ \citep[e.g.,][]{Cover+Thomas05Book}.
$\dot{{\cal I}}$ is called the information flow and represents the change in the mutual information ${\cal M}(T_t:h_t)$ induced by the time evolution of $T$ \citep{Allahverdyan+09JStatMech,Horowitz+Esposito14PhysRevX,Rosinberg+Horowitz2016,Loos+Klapp20NewJPhys}.
Here, ${\cal M}(T_t:h_t)$ is defined as:
\begin{align}
    {\cal M}(T_t:h_t) := \int {\rm d} T_t\,{\rm d} h_t\;p_t(T_t, h_t)\ln \frac{p_t(T_t, h_t)}{p_t(T_t)p_t(h_t)}. \label{eq:def-m}
\end{align}
Mutual information quantifies statistical dependence between variables, which can be interpreted as their correlation \citep[e.g.,][]{Cover+Thomas05Book}.
Thus, $\dot{{\cal I}}$ can be interpreted as the change in the correlation between $T$ and $h$ induced by changes in $T$ alone.
Further discussion of these quantities ($\dot{\cal Q}$, ${\cal S}$, and $\dot{{\cal I}}$) is given in Appendix~A\ref{app-subsec:second-law}.

In the SRO, $\dot{\cal Q}$ is dominant in $\dot{{\cal P}}$ (Section~\ref{sec:results}).
In the remainder of this subsection, we discuss why $\dot{\cal Q}$ is an indicator of irreversibility.
Note that the relative importance of the terms in $\dot{{\cal P}}$ differs from system to system.
For example, when nonstationarity is strong, the time derivative ${\rm d}{\cal S}/{\rm d}t$ may be important \citep[e.g.,][]{Ito+Sagawa15NatComm,Matsumoto+Sagawa2018PRE}.
As another example, in Maxwell's demon (a central problem in stochastic thermodynamics), the sign and magnitude of the information flow $\dot{{\cal I}}$ are essential \citep[e.g.,][]{Sagawa+Ueda13InBook,Yasuda+Kohyama2025JCLI}.

The quantity $\dot{\cal Q}$ in Eq.~(\ref{eq:def-q}) is given by the ratio of the forward and backward transition probabilities of $T$ \citep{Sekimoto10Book,Seifert12RepProgPhys,Ito+Sagawa2016inbook}.
Here we briefly discuss the result; Appendix~A\ref{app-subsec:q} reviews the derivation.
Suppose that the state at time $t$ is $(T_t, h_t)$.
The independence of $\xi^T$ and $\xi^h$ allows the transitions of $T$ and $h$ to be treated separately \citep[the bipartite condition; e.g.,][]{Horowitz+Esposito14PhysRevX}.
When $T_{t+{\rm d}t} = T_t + {\rm d}T$, the corresponding transition density is written as $p(T_{t+{\rm d}t} \mid T_t, h_t)$, and the backward transition density as $p(T_t \mid T_{t+{\rm d}t}, h_t)$.
Then, $\dot{\cal Q}$ is given by:
\begin{align}
    \left\langle \ln \frac{p_t(T_{t+{\rm d}t} \mid T_t, h_t)}{p_t(T_t \mid T_{t+{\rm d}t}, h_t)} \right\rangle &= \frac{2}{\left(\sigma^T\right)^2} \dot{\cal Q} \, {\rm d}t. \label{eq:def-q-alternative}
\end{align}
The more irreversible the transition, the larger $\dot{\cal Q}$.
Here, irreversibility means that the forward transition is more likely than the backward.
\revA{Irreversibility is further discussed in Sections~\ref{sec:tur}\ref{subsec:tur-one-variable}, \ref{sec:results}, and~\ref{sec:discussion}\ref{subsec:interpretation}.}

\subsection{Specific form of the TUR} \label{subsec:tur-specific}

We choose the weight function $g(T,h)$ as follows:
\begin{align}
    g(T,h) = T.
\end{align}
In this case, the mean current becomes
\begin{equation}
    \left \langle {\cal J} \right \rangle = \left\langle T \circ {\rm d}T \right\rangle = \frac{1}{2} \left\langle \frac{{\rm d}T^2}{{\rm d}t} \right\rangle {\rm d}t = \frac{1}{2} \left(\frac{{\rm d}}{{\rm d}t} \left\langle T^2\right\rangle\right) {\rm d}t, \label{eq:mean-current-specific}
\end{equation}
where we use the fact that the Stratonovich product $\circ$ satisfies the chain rule of differentiation \citep[e.g.,][]{Gardiner09Book}.
The current ${\cal J}$ generally represents a net displacement; with $g(T,h)=T$, the mean current $\left \langle {\cal J} \right \rangle$ is proportional to the tendency of $\left\langle T^2\right\rangle$.
Next, ${\rm Var}[{\cal J}]$ becomes (see Appendix~A\ref{app-subsec:tur-specific} for the derivation):
\begin{align}
    {\rm Var}[{\cal J}] &= \left(\sigma^T\right)^2\left\langle T^2 \right\rangle {\rm d}t, \label{eq:tur-variance-specific}
\end{align}
which is proportional to $\left\langle T^2 \right\rangle$ and explicitly contains the squared noise amplitude $\left(\sigma^T\right)^2$.

Thus, the TUR~(\ref{eq:tur-general2}) reduces to:
\begin{align}
    \frac{\dot{\cal P}}{2} &\ge \frac{1}{\left(\sigma^T\right)^2\left\langle T^2 \right\rangle} \left(\frac{1}{2} \frac{{\rm d}}{{\rm d}t} \left\langle T^2 \right\rangle \right)^2.\label{eq:tur-main1}
\end{align}
The left-hand side involves the partial entropy production rate $\dot{\cal P}$ in Eq.~(\ref{eq:def-pdot}), whereas the right-hand side is the precision (Section~\ref{sec:tur}\ref{subsec:tur-general}).
For the normalized squared tendency of $\left\langle T^2 \right\rangle$ to be large, $\dot{\cal P}$ must be sufficiently large.
In particular, if $\dot{\cal P} = 0$, then $\left\langle T^2 \right\rangle$ must be constant in time.
This inequality holds at any time and for any initial condition \citep{Tanogami+2023PRR}, provided that ${\rm Var}[{\cal J}]$ in Eq.~(\ref{eq:tur-variance-specific}) is nonzero.
Since the derivation only assumed $g(T,h)=T$, it did not use the linearity of the SRO (see Appendices~A\ref{app-subsec:tur-general} and~A\ref{app-subsec:tur-specific}).
Thus, an inequality of the same form holds even if nonlinear terms are added to the governing equations, as long as the noise remains additive.

By contrast, the explicit expressions for the terms in $\dot{\cal P}$ [i.e., Eqs.~(\ref{eq:def-q})--(\ref{eq:def-i})] depend on the form of the SRO governing equations.
For example, $\dot{\cal Q}$ is generally given by the ratio of the forward and backward transition probabilities (Section~\ref{sec:tur}\ref{subsec:tur-general}), but its explicit expression follows from the governing equations.
These expressions are given in Appendix~B, as they are needed for the analysis in Section~\ref{sec:results}.
By the linearity of the SRO, $\langle T \rangle$ and $\langle h \rangle$ converge to zero as the influence of the initial conditions decays, so $\left\langle T^2 \right\rangle$ can be identified with the variance of $T$.
Hereafter, we refer to $\left\langle T^2 \right\rangle$ as the variance of $T$ when there is no confusion.

\subsection{\revA{Irreversibility in a one-variable model}} \label{subsec:tur-one-variable}

\revA{To clarify the meaning of irreversibility for SSTA, we consider the one-variable model obtained by setting $F_1 = 0$ in Eq.~(\ref{eq:linear-sro1}) and discarding $h$:
\begin{align}
    \frac{{\rm d} T}{{\rm d} t} = R(t) T + \sigma^T \xi^T, \label{eq:one-variable-model-main}
\end{align}
where we assume $R_0 < 0$.
Without seasonality (i.e., $R_a = 0$ and $R(t) = R_0$), Eq.~(\ref{eq:one-variable-model-main}) has the same form as the stochastic climate model of \citet{Hasselmann1976Tellus} and \citet{Frankignoul+Hasselmann1977Tellus}.
With or without seasonality, the TUR~(\ref{eq:tur-main1}) holds with equality at every time once the influence of the initial condition has decayed (Appendix~C\ref{app-subsec:one-variable}):
\begin{align}
    \frac{\dot{\cal P}}{2} = \frac{1}{\left(\sigma^T\right)^2 \left\langle T^2 \right\rangle} \left( \frac{1}{2}\frac{{\rm d}}{{\rm d}t} \left\langle T^2 \right\rangle \right)^2. \label{eq:tur-equality-one-variable-main}
\end{align}

In the model~(\ref{eq:one-variable-model-main}), $\dot{\cal P}$ measures the distinguishability between the forward and backward (i.e., time-reversed) processes of $T$ \citep[e.g.,][]{Seifert12RepProgPhys}.
The total probability of each process is the product of the probability densities of its initial state and its transition.
The ratio of these total probabilities gives $\dot{\cal P}$ (Appendix~C\ref{app-subsec:one-variable}), whereas $\dot{\cal Q}$ describes the ratio of the transition probabilities alone (Section~\ref{sec:tur}\ref{subsec:tur-general} and Appendix~A\ref{app-subsec:q}).

Without seasonality ($R_a = 0$), the damping by $R_0$ balances the fluctuations driven by the noise.
The variance becomes constant, $\left\langle T^2 \right\rangle = -\left(\sigma^T\right)^2/(2R_0)$, leading to $\dot{\cal P} = 0$ in Eq.~(\ref{eq:tur-equality-one-variable-main}).
Thus, a trajectory of $T$ driven by the noise is as likely as its time-reversed trajectory (Fig.~\ref{fig:one-variable-reversal}a,c).
That is, the forward and time-reversed processes are statistically indistinguishable, and in this sense the evolution of $T$ is reversible.

With seasonality ($R_a \neq 0$), this balance is broken.
$\left\langle T^2 \right\rangle$ cannot instantly follow the seasonal change in $R(t)$.
The resulting increases and decreases in $\left\langle T^2 \right\rangle$ make the forward and time-reversed processes distinguishable.
For example, in the forward process, $\left\langle T^2 \right\rangle$ increases around September, when the damping by $R(t)$ is weakest.
If time were reversed, this evolution would instead show a \emph{decrease} in $\left\langle T^2 \right\rangle$ around September, when the damping is weakest.
Such a decrease is unlikely in the time-reversed process, in which only the seasonal cycle of $R(t)$ is reversed.
This example can also be illustrated with a single trajectory of $T$.
That is, $T$ peaks after and \emph{before} the maximum of $R(t)$ in Fig.~\ref{fig:one-variable-reversal}b and~\ref{fig:one-variable-reversal}d, respectively.
The latter is unlikely, leading to the distinguishability between the forward and time-reversed processes.

\begin{figure}[t]
    \centering
    \noindent\includegraphics[width=14cm]{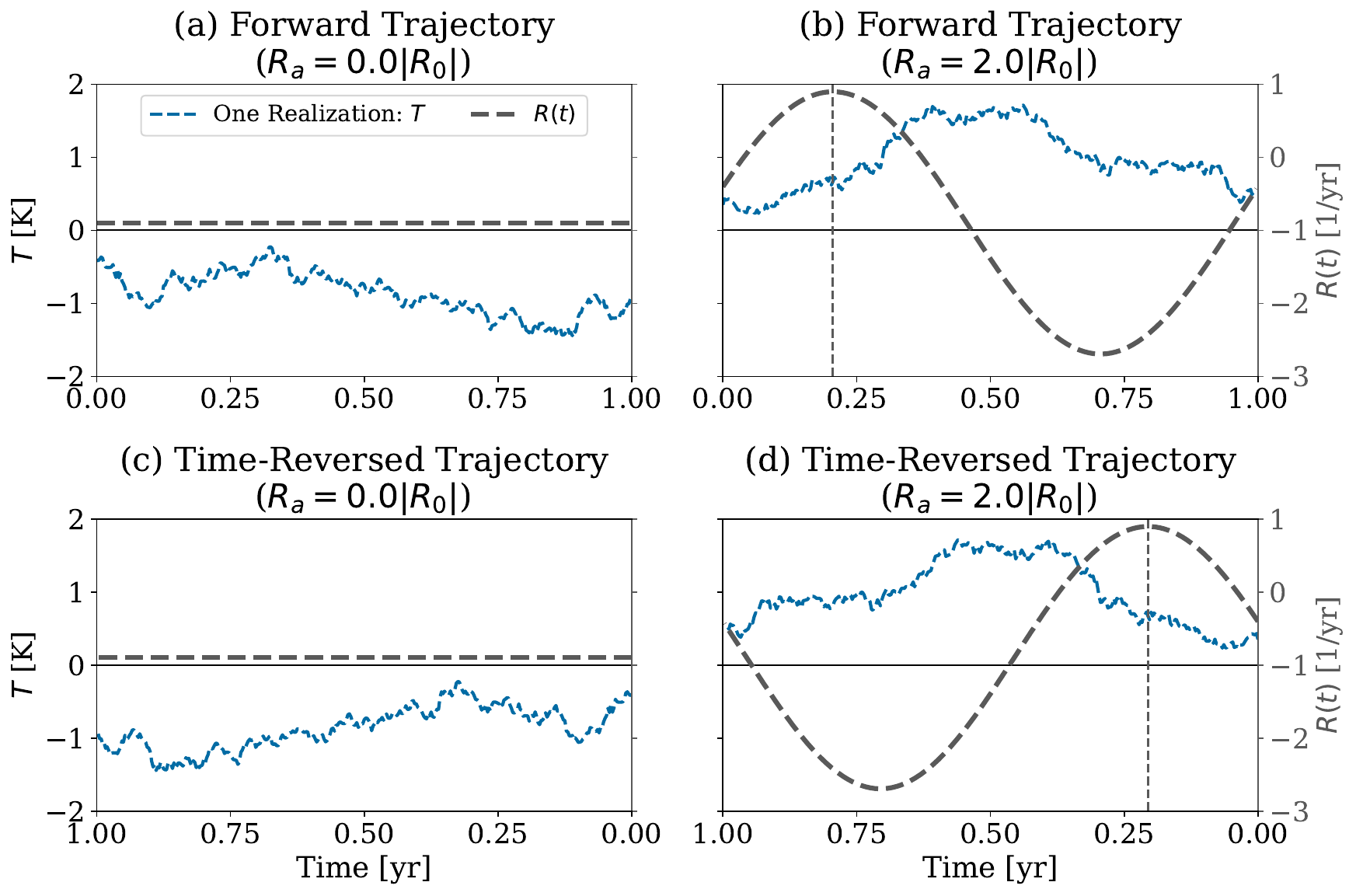}
    \caption{
        \revA{Time series of the one-variable model [Eq.~(\ref{eq:one-variable-model-main})] with the parameters of \citetalias{Kim+An2021JCLI} (Table~\ref{table:sro-parameters}).
        (a),(b) Forward trajectories of $T$ for $R_a = 0$ and $R_a = 2.0 \lvert R_0 \rvert$.
        (c),(d) Time-reversed trajectories of $T$, that is, the trajectories in (a),(b) reversed in time.
        Each panel also shows the growth rate $R(t)$ (right axis), which is constant in (a) and (c).
        Time is measured from 1 July, and the vertical dashed lines in (b) and (d) mark the maximum of $R(t)$ in mid-September, when the damping is weakest.}
    }
    \label{fig:one-variable-reversal}
\end{figure}

In the model~(\ref{eq:one-variable-model-main}), irreversibility is therefore equivalent to the delayed response of the SSTA variance to the seasonal change in $R(t)$.
As an analogy from thermodynamics, a gas compressed slowly by a piston changes quasi-statically and reversibly, whereas under rapid compression, the relaxation of the gas lags behind the piston motion, and entropy is produced.
The rate of this entropy production corresponds to $\dot{\cal P}$ in stochastic thermodynamics \citep[e.g.,][]{Peliti+Pigolotti21Book,Shiraishi23Book}.
In the SRO with both $T$ and $h$ (Section~\ref{sec:results}), $\dot{\cal P}$ is split into the irreversibility associated with this delayed response and that associated with the ENSO cycle (i.e., the recharge and discharge of equatorial heat content $h$).}

\subsection{Limitations in applying the TUR} \label{subsec:tur-limitations}

Although TURs hold broadly \revA{as constraints on fluctuations} \citep[e.g.,][]{Horowitz+Gingrich2020NatPhys}, they have at least four limitations.
First, the TUR~(\ref{eq:tur-main1}) is an inequality, and thus the tightness of the bound is important.
For example, if $\dot{\cal P}$ greatly exceeds the right-hand side of Eq.~(\ref{eq:tur-main1}) over a realistic parameter range, the TUR is trivially satisfied and provides little insight.
Conversely, the TUR is more useful when the bound is tight.

Second, the TUR constrains statistical averages, not individual realizations.
For example, even when the variance of a quantity is strongly suppressed, the probability of observing a large value is generally nonzero.
In this sense, the TUR describes statistical tendencies.

Third, the TUR~(\ref{eq:tur-main1}) is a diagnostic relation and does not describe causality.
It states that a rapid change in $\left\langle T^2 \right\rangle$ requires a sufficiently large $\dot{\cal P}$.
That is, the TUR constrains the time evolution of $\left\langle T^2 \right\rangle$ \revA{but does not determine it.}

Fourth, the TUR~(\ref{eq:tur-main1}) alone cannot determine the sign of the tendency of $\left\langle T^2 \right\rangle$, because it involves only the square of ${\rm d}\left\langle T^2 \right\rangle/{\rm d}t$.
In Section~\ref{sec:results}, noting that $\left\langle T^2 \right\rangle$ is an indicator of uncertainty, we use the Shannon entropy in Eq.~(\ref{eq:def-s}) to diagnose the sign of the tendency.

\FloatBarrier

\section{Results} \label{sec:results}

\revA{In the SRO, the TUR gives an upper bound on the tendency of $\left\langle T^2 \right\rangle$.}
Here, we show that with increasing annual-cycle amplitude $R_a$, seasonality emerges in the partial entropy production rate $\dot{\cal P}$, dominated by the irreversibility indicator $\dot{\cal Q}$.
The seasonality of $\dot{\cal Q}$ \revA{is associated mainly with} air--sea interaction in the SRO.
These results suggest that the seasonal variation of $\left\langle T^2 \right\rangle$ is constrained by air--sea irreversibility.

\begin{figure}[t]
    \centering
    \noindent\includegraphics[width=16.45cm]{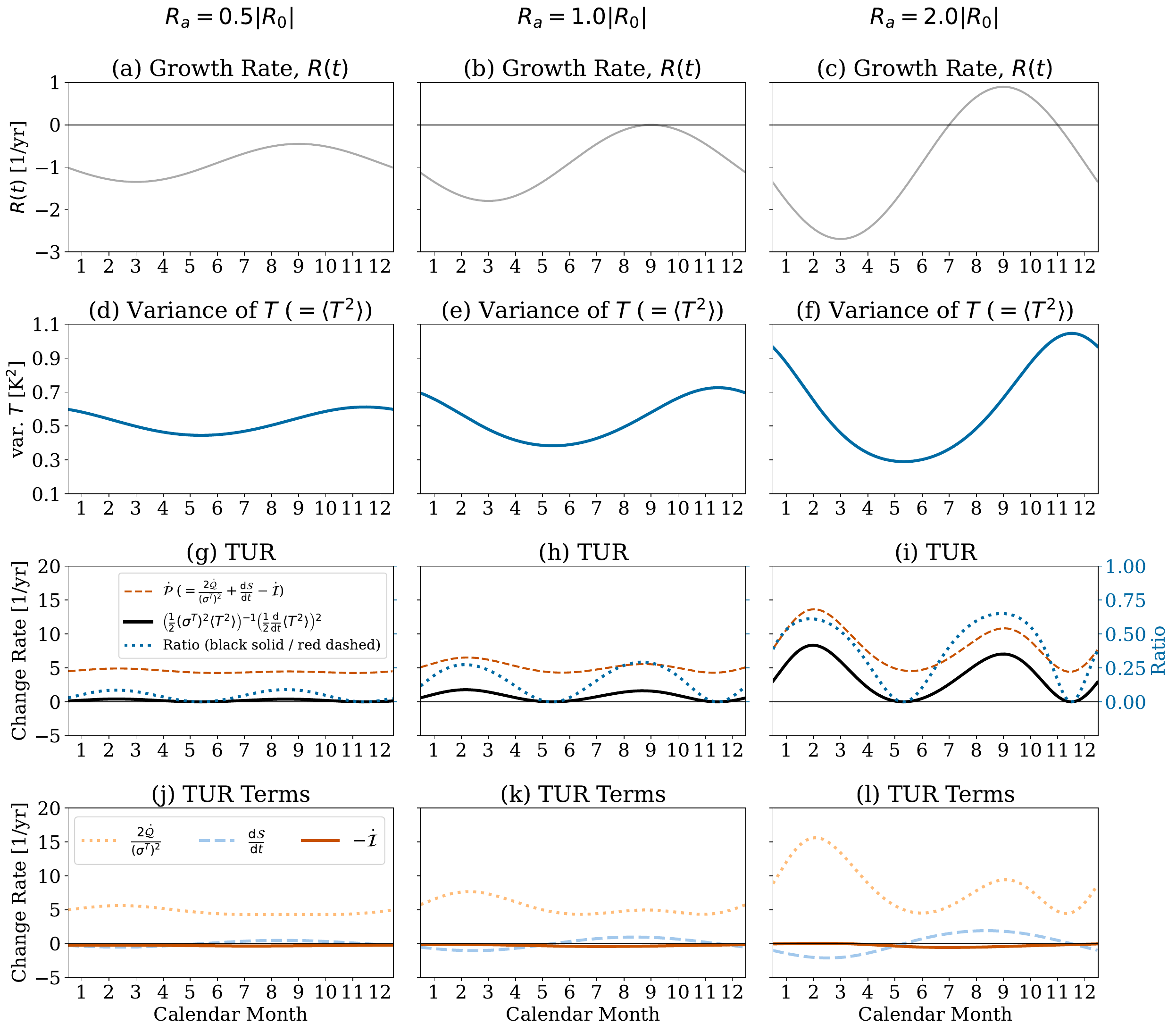}
    \caption{
        Application of the TUR~(\ref{eq:tur-main2}) to $T$.
        (a)--(c) Growth rate $R(t)$.
        (d)--(f) Variance of $T$, $\left\langle T^2 \right\rangle$.
        (g)--(i) Normalized square of the tendency of $\left\langle T^2 \right\rangle$, its TUR upper bound $\dot{\cal P}$, \revA{and the ratio of the former to the latter (right axis).}
        (j)--(l) Components of $\dot{\cal P}$ [Eq.~(\ref{eq:def-pdot})], namely ${2\dot{\cal Q}}/{\left(\sigma^T\right)^2}$, ${{\rm d} {\cal S}}/{{\rm d} t}$, and $-\dot{{\cal I}}$.
        These TUR quantities were obtained from the means and covariance matrix of $T$ and $h$ (Appendix~B), integrated for 200 years; the final year is shown.
    }
    \label{fig:tur-KA21}
\end{figure}

Figure~\ref{fig:tur-KA21} shows the annual cycles of the growth rate $R(t)$, the variance $\left\langle T^2 \right\rangle$, and the TUR terms.
We numerically integrated the evolution equations for the means and covariance matrix of $T$ and $h$, and then computed the TUR terms from these quantities (Appendix~B).
After spin-up, the system is nearly cyclostationary.

As $R_a$ increases, the annual-cycle amplitude of $\left\langle T^2 \right\rangle$ also increases (Fig.~\ref{fig:tur-KA21}a--f), consistent with \citetalias{Kim+An2021JCLI} (their Fig.~A1) and indicating stronger phase locking.
The black solid lines in Fig.~\ref{fig:tur-KA21}g--i show the normalized squared tendency of $\left\langle T^2 \right\rangle$ on the right-hand side of the TUR:
\begin{align}
    \dot{\cal P} &\ge \frac{\left(\frac{1}{2} \frac{{\rm d}}{{\rm d}t}  \left\langle T^2\right\rangle \right)^2}{\frac{1}{2}{\left(\sigma^T\right)^2\left\langle T^2 \right\rangle}}.\label{eq:tur-main2}
\end{align}
The variance $\left\langle T^2 \right\rangle$ increases from July to November and decreases from December to April (Fig.~\ref{fig:tur-KA21}d--f).
The square of its tendency thus attains maxima around September and February (black solid lines; a wavenumber-2 structure in Fig.~\ref{fig:tur-KA21}g--i).
The partial entropy production rate $\dot{\cal P}$ closely tracks this structure (red dashed lines).
$\dot{\cal P}$ consists of three terms [Eq.~(\ref{eq:def-pdot})], whose time series are shown in Fig.~\ref{fig:tur-KA21}j--l.
The irreversibility indicator ${2\dot{\cal Q}}/{\left(\sigma^T\right)^2}$ dominates, whereas the other two terms, ${{\rm d} {\cal S}}/{{\rm d} t}$ and $\dot{{\cal I}}$, contribute little.

\FloatBarrier

The TUR~(\ref{eq:tur-main2}) means that net changes in $\left\langle T^2 \right\rangle$, whether increases or decreases, are constrained by $\dot{\cal P}$.
Noise induces both net changes and fluctuations, which appear in the numerator and denominator of the right-hand side, respectively (Section~\ref{sec:tur}\ref{subsec:tur-general}).
\revA{$\dot{\cal P}$ must be large enough for a net change to overcome the fluctuations \citep[e.g.,][]{Horowitz+Gingrich2020NatPhys,Tanogami+2023PRR}.
When $\dot{\cal P}$ is not large enough, this constraint limits the net change.}
The minimum of $\dot{\cal P}$ varies little with $R_a$, whereas its maximum increases with $R_a$ (Fig.~\ref{fig:tur-KA21}g--i).
Around September and February, $\dot{\cal P}$ becomes large, relaxing the TUR constraint and permitting a net change in $\left\langle T^2 \right\rangle$.
The variance $\left\langle T^2 \right\rangle$ then reaches its maximum around November, corresponding to the concentration of El Ni\~no peaks (discussed in more detail below).
Conversely, for small $R_a$, $\dot{\cal P}$ is nearly uniform across seasons, and the seasonality of $\left\langle T^2 \right\rangle$ is weak.
In this case, fluctuations become dominant, and $T$ can peak in any calendar month.

The leading term of $\dot{\cal P}$, namely ${2\dot{\cal Q}}/{\left(\sigma^T\right)^2}$, has maxima around September and February mainly because $\lvert R(t) \rvert$ becomes large near these times: $R(t)$ attains its maximum around September and its minimum around March (Fig.~\ref{fig:tur-KA21}a--c).
Intuitively, when $R(t)$ is minimal (that is, when it is most negative), $R(t)T$ acts as damping of $T$ in Eq.~(\ref{eq:linear-sro1}), increasing its irreversibility.
Conversely, when $R(t)$ is maximal, $R(t)T$ acts as amplification of $T$, also increasing its irreversibility.
\revA{In both cases, $\left\langle T^2 \right\rangle$ lags behind $R(t)$ and is far from the balance between damping and noise (Section~\ref{sec:tur}\ref{subsec:tur-one-variable}).}
Both the minimum and the maximum of $R(t)$ therefore correspond to maxima of irreversibility, together producing the wavenumber-2 structure of ${2\dot{\cal Q}}/{\left(\sigma^T\right)^2}$.

The next component of $\dot{\cal P}$, the Shannon entropy tendency ${\rm d}{\cal S}/{\rm d}t$, diagnoses whether $\left\langle T^2 \right\rangle$ is increasing or decreasing, because ${\cal S}$ is a monotonic function of $\left\langle T^2 \right\rangle$ in the SRO by Gaussianity (Appendix~B).
This sign information cannot be obtained from the TUR alone (Section~\ref{sec:tur}\ref{subsec:tur-limitations}).
\revA{Specifically, ${\rm d}{\cal S}/{\rm d}t$} takes its maximum in August--September, when $\left\langle T^2 \right\rangle$ is increasing; ${\cal S}$ itself is then maximal in November--December, when $\left\langle T^2 \right\rangle$ is large.

The final component of $\dot{\cal P}$, the information flow $\dot{{\cal I}}$ in Eq.~(\ref{eq:def-i}), is the smallest and almost negligible.
The information flow represents the change in the correlation between $T$ and $h$ induced by changes in $T$ (Section~\ref{sec:tur}\ref{subsec:tur-general}).
In the SRO, $T$ and $h$ fluctuate under noise but tend to be out of phase, with $h$ leading $T$ (Section~\ref{sec:sro}).
Although the \revA{lagged} correlation itself is strong, its tendency is small, likely keeping $\dot{\cal I}$ small.

\revA{To quantify the tightness of the TUR bound (Section~\ref{sec:tur}\ref{subsec:tur-limitations}), Fig.~\ref{fig:tur-KA21}g--i also shows the ratio (blue dotted lines) of the right-hand side of Eq.~(\ref{eq:tur-main2}) to its left-hand side $\dot{\cal P}$.
This ratio is largest around September and February, when $\dot{\cal P}$ (red dashed lines) is also large.
The maximum ratio increases with $R_a$: 0.09, 0.29, and 0.65 for $R_a = 0.5$, 1.0, and $2.0 \lvert R_0 \rvert$, respectively, indicating that the bound becomes tighter as phase locking strengthens.
With the other parameter sets in Table~\ref{table:sro-parameters}, those of \citet{Vialard+2025RevGeophys} and \citet{Han+2026JCLI}, the maximum ratio is about 0.73, larger than that for \citetalias{Kim+An2021JCLI} (0.65).
In all cases, however, the ratio remains below 1 throughout the year, unlike in the one-variable model, in which the TUR holds with equality and the ratio is always 1 (Section~\ref{sec:tur}\ref{subsec:tur-one-variable}).

In the SRO, this gap between the two sides of the TUR~(\ref{eq:tur-main2}) is attributed to the ENSO cycle, that is, the recharge and discharge of the equatorial heat content $h$.
Indeed, $\dot{\cal P}$ splits exactly into two parts (Appendix~C\ref{app-subsec:two-variable}):
\begin{align}
    \dot{\cal P} = \underbrace{\frac{\left\langle T \circ \frac{{\rm d}T}{{\rm d}t} \right\rangle^2}{\frac{1}{2}\left(\sigma^T\right)^2\left\langle T^2 \right\rangle}}_{=:\dot{\cal P}_{\rm var}} + \underbrace{\frac{\left\langle h_\perp \circ \frac{{\rm d}T}{{\rm d}t} \right\rangle^2}{\frac{1}{2}\left(\sigma^T\right)^2\left\langle h_\perp^2 \right\rangle}}_{=:\dot{\cal P}_{\rm cyc}}, \label{eq:split-main}
\end{align}
where $h_\perp$ is the part of $h$ uncorrelated with $T$, and the subscripts ``var'' and ``cyc'' refer to the variance and the ENSO cycle, respectively.
The first term, $\dot{\cal P}_{\rm var}$, is the right-hand side of the TUR~(\ref{eq:tur-main2}) because $\left\langle T \circ {\rm d}T/{\rm d}t \right\rangle = (1/2)\,{\rm d}\left\langle T^2 \right\rangle/{\rm d}t$ [Eq.~(\ref{eq:mean-current-specific})].
Thus, the ratio in Fig.~\ref{fig:tur-KA21}g--i is $\dot{\cal P}_{\rm var}/\dot{\cal P}$.
This ratio is smaller than 1 because of the second term, $\dot{\cal P}_{\rm cyc}$, which represents the deviation of the TUR from equality.
As in the one-variable model (Section~\ref{sec:tur}\ref{subsec:tur-one-variable}), $\dot{\cal P}_{\rm var}$ is associated with the delayed response of $\left\langle T^2 \right\rangle$ to the seasonal change in $R(t)$.
In contrast, $\dot{\cal P}_{\rm cyc}$ is associated with the ENSO cycle (Appendix~C\ref{app-subsec:pdot-osc-explicit}) and remains positive even for $R_a = 0$ (i.e., even when $\dot{\cal P}_{\rm var} = 0$).
Moreover, the seasonal variation of $\dot{\cal P}_{\rm cyc}$ is weaker than that of $\dot{\cal P}_{\rm var}$, likely because $\dot{\cal P}_{\rm cyc}$ contains no $R(t)$ explicitly (Appendix~C\ref{app-subsec:pdot-osc-explicit}).

Since $\dot{\cal P}$ is dominated by the irreversibility indicator $\dot{\cal Q}$, these results suggest that $\dot{\cal Q}$ also contains contributions from the seasonal change in $R(t)$ and from the ENSO cycle.
The latter reflects the preferred temporal direction of the recharge--discharge cycle, with $h$ leading $T$ (Section~\ref{sec:sro}).
This temporal direction makes the forward and time-reversed processes distinguishable.
For simplicity, we first consider the SRO without noise, in which $h$ leads $T$ in the damped oscillation, so the state $(T,h)$ \emph{rotates} in a fixed direction in the state space.
In the forward trajectory, the heat content is recharged before El Ni\~no develops.
In the time-reversed trajectory, the rotation would be reversed, and El Ni\~no would develop before the recharge, which does not occur under Eqs.~(\ref{eq:linear-sro1}) and~(\ref{eq:linear-sro2}) without noise.
With noise, such a reversed evolution can occur, but it is unlikely.
The two processes thus remain statistically distinguishable, which corresponds to positive entropy production.
We further discuss the decomposition of $\dot{\cal Q}$ later in this section.}

\revA{Previous studies show that linear stochastic models with constant coefficients are irreversible only when there is a mean rotation in the state space \citep{Weiss07,Benincasa26}.
Such irreversibility has been identified in observed tropical SST and ENSO indices \citep{Weiss09,Weiss+20}.
Our result is consistent with these studies.
The SRO with constant coefficients (i.e., $R_a = 0$) has $\dot{\cal P} = \dot{\cal P}_{\rm cyc} > 0$, which is attributed to the irreversible rotation in the $(T,h)$ space.
In contrast to these previous studies, we split $\dot{\cal P}$ exactly as in Eq.~(\ref{eq:split-main}) for the SRO with seasonally varying $R(t)$ ($R_a \neq 0$).
This split reveals that irreversibility is separated into the part associated with the seasonal variation of $\left\langle T^2 \right\rangle$, which reflects phase locking, and the part associated with the rotation in the state space, which corresponds to the ENSO (or recharge--discharge) cycle.}

We next examine the calendar months in which El Ni\~no peaks tend to occur in the SRO.
We follow the procedure of \citetalias{Kim+An2021JCLI}.
We varied $R_a$ from 0 to $2.5 \lvert R_0 \rvert$ and, for each value, integrated a single ensemble member for 50,000 years.
For each time series, we computed the standard deviation of $T$ and used it as the El Ni\~no threshold.
We then applied a 5-month running mean to $T$ \revA{to smooth out intraseasonal variations} and defined an El Ni\~no event as a time interval during which $T$ continuously exceeds this threshold for at least 6 months \revA{\citep{Trenberth97}.}
We extracted the local maxima of $T$ within each event and counted them for each calendar month.
Multiple maxima within a single event were counted separately if they were at least 5 months apart.
The results are not sensitive to the integration length or running-mean window.
Similar peak distributions were obtained with a 10,000-yr integration, with different running-mean windows, and when only the global maximum of $T$ in each event was counted.

\begin{figure}[t]
    \centering
    \noindent\includegraphics[width=16.45cm]{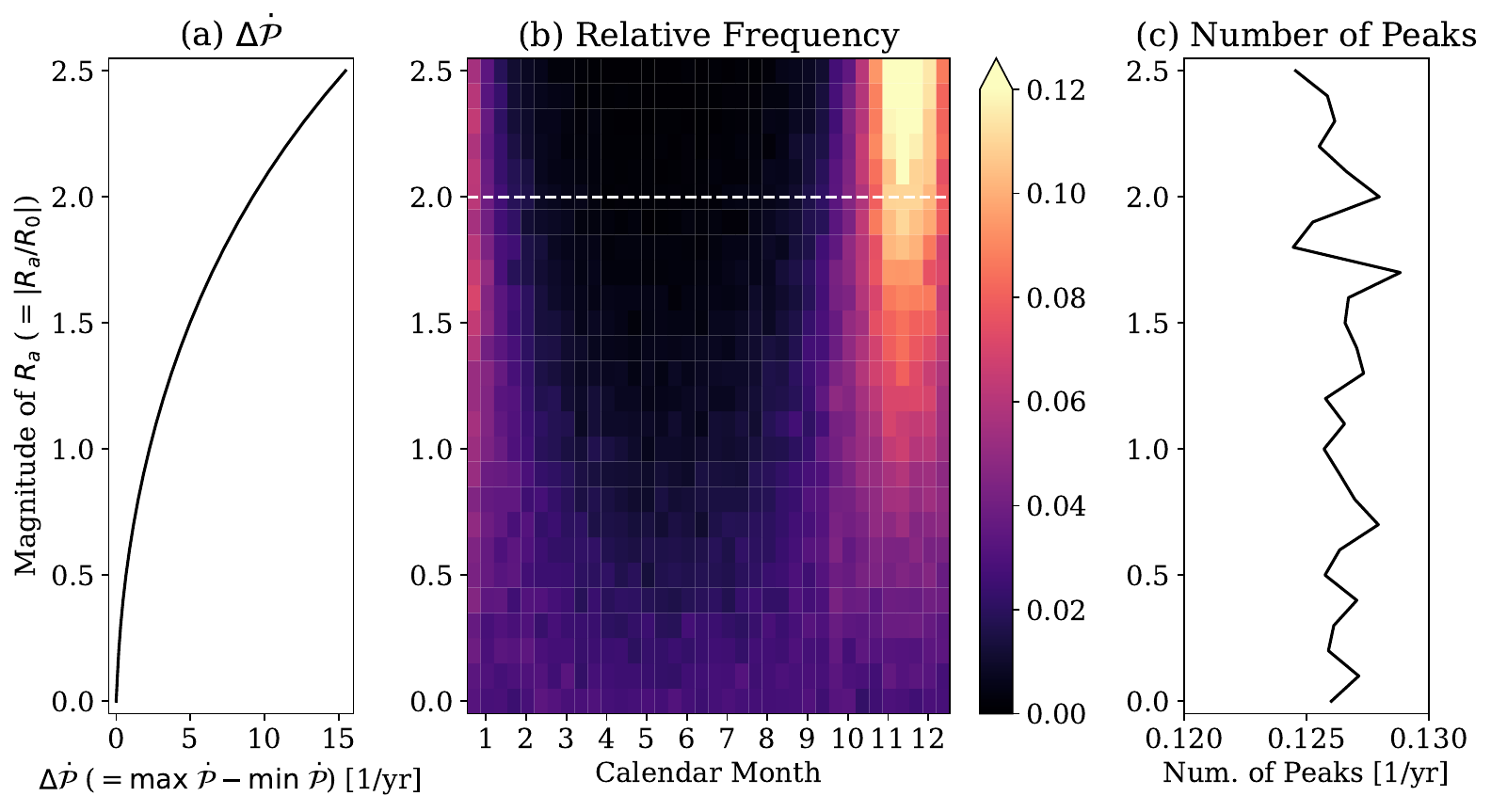}
    \caption{
        Distributions of $T$ peaks in the SRO.
        (a) $\Delta \dot{\cal P}$, the range of $\dot{\cal P}$.
        (b) Relative frequency of peaks across calendar months.
        For each $R_a$, the frequencies sum to unity.
        The white horizontal dashed line marks $R_a = 2.0 \lvert R_0 \rvert$, the value for observed ENSO \citepalias{Kim+An2021JCLI}.
        (c) Annual-mean number of peaks.
    }
    \label{fig:peak-seasonality-KA21}
\end{figure}

Figure~\ref{fig:peak-seasonality-KA21} shows the peak distributions.
As $R_a$ increases, the peaks become concentrated in November--December (Fig.~\ref{fig:peak-seasonality-KA21}b), consistent with \citetalias{Kim+An2021JCLI}.
The horizontal dashed line marks $R_a = 2.0 \lvert R_0 \rvert$, the value corresponding to observed ENSO \citepalias{Kim+An2021JCLI}.
The number of peaks is approximately 0.125 per year (Fig.~\ref{fig:peak-seasonality-KA21}c) and is nearly independent of $R_a$: their total changes little while their seasonal concentration strengthens.
We characterize this concentration by the difference $\Delta \dot{\cal P}$ (Fig.~\ref{fig:peak-seasonality-KA21}a), defined as
\begin{equation}
    \Delta \dot{\cal P} := \max \;\dot{\cal P} - \min \;\dot{\cal P}.
\end{equation}
\revA{$\min \;\dot{\cal P}$ nearly equals $\dot{\cal P}_{\rm cyc}$ because $\dot{\cal P}_{\rm var}$ vanishes at the extrema of $\left\langle T^2 \right\rangle$ [Eq.~(\ref{eq:split-main})].
Moreover, as discussed above, the seasonal variation of $\dot{\cal P}_{\rm var}$ is stronger than that of $\dot{\cal P}_{\rm cyc}$, so $\Delta \dot{\cal P}$ mainly reflects the range of $\dot{\cal P}_{\rm var}$, which is associated with the seasonal change in $R(t)$.}
Larger $\Delta \dot{\cal P}$ corresponds to stronger seasonality of $\left\langle T^2 \right\rangle$ and to stronger phase locking of El Ni\~no in the SRO (Fig.~\ref{fig:peak-seasonality-KA21}a,b).
\revA{This correspondence broadly holds even when $R_a$ is fixed and the other parameters are varied (Appendix~E).}
\revA{Thus, $\Delta \dot{\cal P}$ can relate the TUR~(\ref{eq:tur-main2}) to the timing of El Ni\~no peaks and is regarded as a diagnostic of phase-locking strength, analogous to the seasonal energy index (SEI) of \citetalias{Kim+An2021JCLI} (Section~\ref{sec:introduction}).}

\FloatBarrier

In preparation for the interpretation \revA{for real ENSO} in Section~\ref{sec:discussion}\ref{subsec:interpretation}, we discuss the cause of the wavenumber-2 structure based on the SRO expression for ${2\dot{\cal Q}}/{\left(\sigma^T\right)^2}$:
\begin{align}
    \frac{2}{\left(\sigma^T\right)^2}\dot{\cal Q} &= \underbrace{\frac{2}{\left(\sigma^T\right)^2} R(t)^2 \left\langle T^2 \right\rangle}_{=A} + \underbrace{\frac{2}{\left(\sigma^T\right)^2} {2 R(t) F_1} \left\langle T h \right\rangle}_{=B} + \underbrace{\frac{2}{\left(\sigma^T\right)^2} \left(F_1\right)^2 \left\langle h^2 \right\rangle}_{=C} + \underbrace{R(t)}_{=D}.\label{eq:decompose-q-a-to-d}
\end{align}
See Appendix~B for the derivation.
The terms $A$, $B$, and $C$ arise from expanding $\langle \left[R(t) T+F_1 h\right]^2\rangle$.
The remaining term $D$ comes from the nonsmoothness of $T$ and is introduced as a correction term through the Stratonovich product \citep[e.g.,][]{Shiraishi23Book}.

\begin{figure}[t]
    \centering
    \noindent\includegraphics[width=16.45cm]{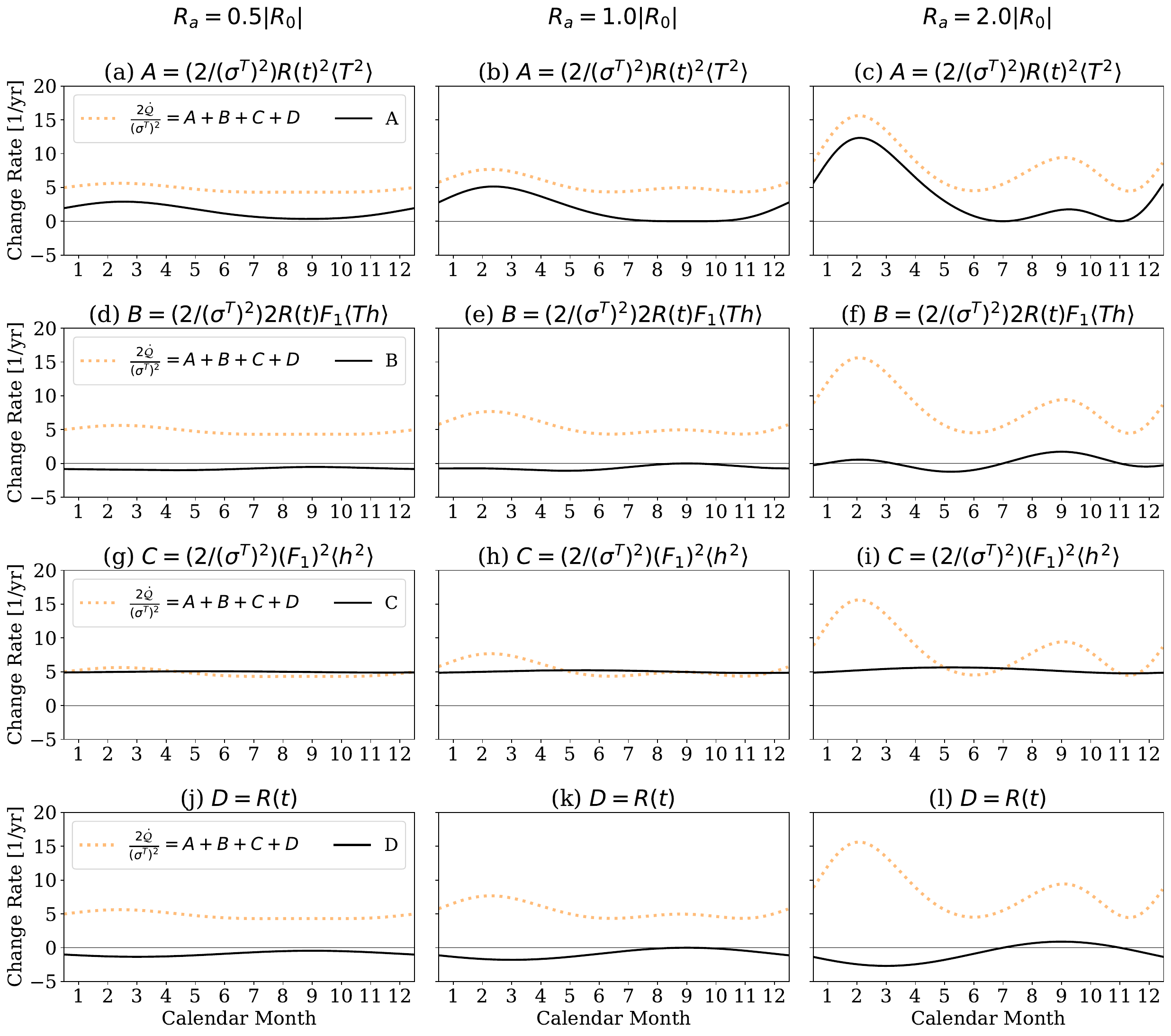}
    \caption{
        Time series of the terms $A$, $B$, $C$, and $D$ of the irreversibility indicator ${2\dot{\cal Q}}/{\left(\sigma^T\right)^2}$ in Eq.~(\ref{eq:decompose-q-a-to-d}).
        \revA{Each panel shows the corresponding term ($A$, $B$, $C$, or $D$) together with ${2\dot{\cal Q}}/{\left(\sigma^T\right)^2}$.}
        These terms were computed from the same simulation as in Fig.~\ref{fig:tur-KA21}; see Appendix~B for the derivation of Eq.~(\ref{eq:decompose-q-a-to-d}).
    }
    \label{fig:decomposition-q-KA21}
\end{figure}

Figure~\ref{fig:decomposition-q-KA21} shows the time series of $A$--$D$.
As $R_a$ increases, the seasonality of $A$--$D$ strengthens, consistent with that of ${2\dot{\cal Q}}/{\left(\sigma^T\right)^2}$ (Fig.~\ref{fig:tur-KA21}j--l).
The correction term $D$ is not directly associated with physical processes.
We therefore focus on $A$, $B$, and $C$.

The seasonality of ${2\dot{\cal Q}}/{\left(\sigma^T\right)^2}$ is produced mainly by $A$ and $B$ (Fig.~\ref{fig:decomposition-q-KA21}a--f), whose relative contributions depend on the choice of $h$.
As in the preceding discussion, we take $h$ to represent the heat content over the entire equatorial Pacific (Section~\ref{sec:sro}); the case of the western equatorial Pacific is discussed in Appendix~D.
The term $A$ has a wavenumber-2 structure originating from $R(t)^2$ [Eq.~(\ref{eq:decompose-q-a-to-d}) and Fig.~\ref{fig:decomposition-q-KA21}a--c].
The growth rate $R(t)$ attains its extrema around September and March, with a larger absolute value in March.
Thus, $A$ peaks in both seasons and is largest in late winter, although the precise timing and amplitude are modulated by the variance $\langle T^2 \rangle$.
The term $B$ has a wavenumber-2 structure originating from the product $R(t)\left\langle T h \right\rangle$ [Eq.~(\ref{eq:decompose-q-a-to-d}) and Fig.~\ref{fig:decomposition-q-KA21}d--f].
The covariance $\left\langle T h \right\rangle$ is largest around August and smallest around early March, approximately in phase with $R(t)$ (Figs.~\ref{fig:annual-cycle-growth-rate-KA21} and~\ref{fig:sro-time-series-KA21}j).

\FloatBarrier

We interpret $A+B$ as the seasonal variation of irreversibility associated with air--sea interaction in the SRO.
The growth rate $R(t)$ includes the Bjerknes feedback (Section~\ref{sec:discussion}\ref{subsec:interpretation}), and $h$ and $T$ interact through atmospheric processes such as wind stress \citep{Jin+2020inbook,Vialard+2025RevGeophys}.
\revA{The terms $A$, $B$, and $D$ contain $R(t)$ explicitly and thus correspond mainly to $\dot{\cal P}_{\rm var}$ [Eq.~(\ref{eq:split-main})], in which $R(t)$ also appears explicitly (Appendix~C\ref{app-subsec:pdot-osc-explicit}).}

In the SRO, this air--sea irreversibility \revA{has a stochastic origin.}
The SRO is driven entirely by stochastic noise (Section~\ref{sec:sro}).
The growth rate $R(t)$ acts as a ``noise rectifier'' that selectively amplifies or damps noise-driven fluctuations depending on the season.
Although each ensemble member fluctuates randomly, an ensemble-wide seasonal bias toward increasing \revA{(or decreasing)} $\left\langle T^2 \right\rangle$ emerges only through this rectification.
\revA{Since $\left\langle T^2 \right\rangle$ cannot instantly follow this rectification, the bias lags behind $R(t)$, making the forward and time-reversed processes distinguishable (Section~\ref{sec:tur}\ref{subsec:tur-one-variable}).}
\revA{The TUR~(\ref{eq:tur-main2}) permits this bias only when the irreversibility indicator $\dot{\cal Q}$ is sufficiently large.}

Finally, the term $C$ in ${2\dot{\cal Q}}/{\left(\sigma^T\right)^2}$ involves the variance $\left\langle h^2 \right\rangle$ of the heat content [Eq.~(\ref{eq:decompose-q-a-to-d})].
Although $\left\langle h^2 \right\rangle$ itself peaks around May--June (Fig.~\ref{fig:sro-time-series-KA21}h), $C$ contributes little to the seasonality of ${2\dot{\cal Q}}/{\left(\sigma^T\right)^2}$ (Fig.~\ref{fig:decomposition-q-KA21}g--i).
Thus, a sufficient heat content alone does not guarantee the strong irreversibility required for rapid growth of $\left\langle T^2 \right\rangle$.
\revA{Indeed, $C$ corresponds to $\dot{\cal P}_{\rm cyc}$, which is associated with the ENSO cycle (Appendix~C\ref{app-subsec:pdot-osc-explicit}).}
An energy-based perspective relies on heat content, whereas our entropy-based perspective focuses on irreversibility.
\revA{The discussion here suggests that the two are related.}
These viewpoints are discussed further in Section~\ref{sec:discussion}\ref{subsec:interpretation}.

\section{Discussion} \label{sec:discussion}

\revA{TURs are generally regarded as inequalities that constrain fluctuations by entropy production \citep[e.g.,][]{Horowitz+Gingrich2020NatPhys}.
In the SRO, irreversibility quantified by the partial entropy production rate $\dot{\cal P}$ also constrains the temporal evolution of the SSTA variance $\langle T^2 \rangle$, a measure of fluctuations (Section~\ref{sec:results}).}

\revA{The TUR can be regarded as a tool for interpreting the solution of the SRO, in which $R(t)$ drives the seasonality.
This is analogous to using the conservation laws of momentum and energy, which follow from the equations of motion, to interpret their solutions.
Unlike these conservation laws, the TUR is an inequality.
This inequality holds even for nonlinear systems that cannot be solved exactly (Sections~\ref{sec:tur}\ref{subsec:tur-general} and~\ref{sec:tur}\ref{subsec:tur-specific}).
Such generality opens the possibility of interpreting more realistic ENSO models in terms of irreversibility.
We discuss the implications of the SRO results for real ENSO (Section~\ref{sec:discussion}\ref{subsec:interpretation}) and then outline such future applications (Section~\ref{sec:discussion}\ref{subsec:future-work}).}

\subsection{\revA{Implications for real ENSO}} \label{subsec:interpretation}

The growth rate $R(t)$ can be computed from climatology and characterizes how anomalies interact with the climatological background state.
The temporal variation of $R(t)T$ can be decomposed into multiple physical processes \citep{Stein+2010JCLI,Jin+2020inbook,Chen+Jin2022GRL,Xing+2024AAS}.
To our knowledge, no consensus has yet been reached on which process is dominant, but the zonal advective feedback (ZA), for example, has been identified as a major term \citep{Chen+Jin2022GRL,Xing+2024AAS}.
ZA represents the advection of the climatological SST by anomalous zonal currents.
If the anomalous zonal current is driven mainly by wind stress, ZA is attributed to the Bjerknes feedback.
The seasonality of ZA has been linked to the seasonal variation of the air--sea coupling coefficient \citep{Chen+Jin2022GRL} and to the seasonal migration of the Intertropical Convergence Zone \citep[ITCZ;][]{Xing+2024AAS}.

\begin{figure}[t]
    \centering
    \noindent\includegraphics[width=16.0cm]{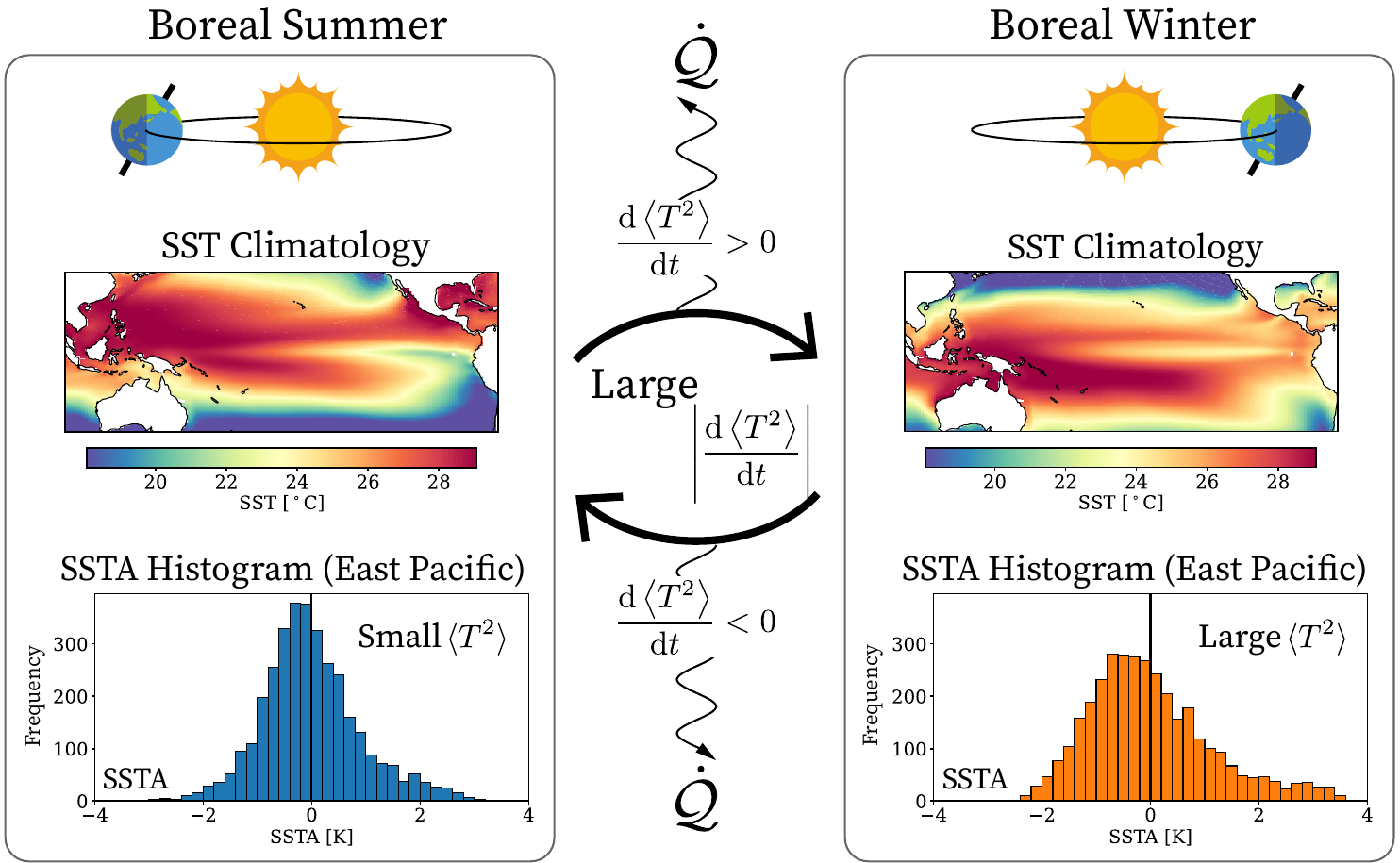}
    \caption{
        Schematic of the irreversibility constraint on ENSO phase locking.
        In boreal summer (left), the SSTA distribution in the eastern Pacific is narrow, corresponding to small variance $\left\langle T^2 \right\rangle$.
        In boreal winter (right), the distribution broadens and $\left\langle T^2 \right\rangle$ is large.
        For the variance to grow or decay between seasons, the TUR requires the irreversibility indicator $\dot{\cal Q}$ (wavy arrows) to be sufficiently large.
        The SST climatology maps and SSTA histograms are computed from the CESM Large Ensemble \citep{Kay+2015BAMS,delaBeaujardiere+2019CESM}.
    }
    \label{fig:schematic}
\end{figure}

\revA{For real ENSO, this seasonal cycle of air--sea coupling is ultimately driven by insolation (Fig.~\ref{fig:schematic}) and may induce a delayed response of the SSTA variance.
This delayed response then corresponds to the ENSO seasonality and can be interpreted as irreversibility, as suggested by the one-variable model (Section~\ref{sec:tur}\ref{subsec:tur-one-variable}).
This irreversibility imposes an entropic constraint.
Entropy quantifies the possibility of a change, whereas the change itself is driven by energy.
The energetic and entropic perspectives are related but distinct, like the first and second laws of thermodynamics \citep[e.g.,][]{Sekimoto10Book,Shiraishi23Book}.
Indeed, the thermal energy of the ocean is abundant but may not be freely available to drive motions in the climate system.
We thus hypothesize that real ENSO is constrained by entropy.}

\revA{Before discussing this hypothesis further, we summarize the energetic and entropic perspectives in the SRO.}
\citetalias{Kim+An2021JCLI} adopted an energetic perspective and showed that the seasonal energy index (SEI) peaks in boreal winter, providing the energy needed for the SSTA variance to grow.
In contrast, we adopt an entropic perspective and find that $\dot{\cal P}$ peaks in boreal autumn and late winter, relaxing the entropy bound on the tendency of $\left\langle T^2 \right\rangle$.
\revA{This $\dot{\cal P}$ is also related to the heat content, whose variance $\left\langle h^2 \right\rangle$ measures oceanic thermal energy.
This variance contributes little to the seasonality of $\dot{\cal P}$ but corresponds to $\dot{\cal P}_{\rm cyc}$, that is, the irreversibility associated with the ENSO cycle (Section~\ref{sec:results}).
This contrast suggests that the oceanic thermal energy is associated with the ENSO cycle, whereas the seasonal variation of irreversibility is associated with phase locking.}

\revA{Beyond the SRO, the relation between energy and entropy has not been established for real ENSO.
In nonequilibrium physics, this relation becomes explicit when the noise source is a heat bath.
The irreversibility indicator $\dot{\cal Q}$, which dominates $\dot{\cal P}$ in the SRO, then equals the energy dissipated as heat \citep{Sekimoto10Book,Seifert12RepProgPhys}.}
It is unclear, however, whether the \revA{small-scale motions acting on ENSO (modeled as noise)} can be interpreted as \revA{thermal noise from} a heat bath, so this dissipated-energy interpretation may not directly apply.
At present, $\dot{\cal Q}$ is understood as an indicator of irreversibility, given by the ratio of the forward and backward transition probabilities \revA{\citep{Sekimoto10Book,Seifert12RepProgPhys}.}
Even so, air--sea interaction typically involves dissipation \citep{Yan+2004GRL,Bannon+2018JGR}, and thus $\dot{\cal Q}$ may also correspond to dissipated energy.

\revA{For real ENSO, if} $\dot{\cal Q}$ were assumed to correspond to some kind of dissipated energy, then our results would be interpreted as follows (Fig.~\ref{fig:schematic}).
$\dot{\cal Q}$ would represent heat dissipated from the subsystem (the equatorial Pacific) to its environment (the rest of the climate system); the equatorial Pacific would then be energetically open.
During the growth phase of ENSO, that is, when the variance $\left\langle T^2 \right\rangle$ increases \revA{from boreal summer to winter,} the dissipation generated by irreversible motions would have to be exported from the equatorial Pacific (upper wavy arrow in Fig.~\ref{fig:schematic}).
During the decay phase, when $\left\langle T^2 \right\rangle$ decreases \revA{from boreal winter to summer,} an analogous export constraint would apply (lower wavy arrow).
However, energy is supplied to the subsystem during growth but released during decay, so the dissipation may differ between the two phases.
Such export could proceed, for example, horizontally to the extratropics via teleconnections, to the other tropical basins (the Indian and Atlantic Oceans) via interbasin interactions, or vertically to space via atmospheric radiation.
\revA{ENSO growth and decay would then be constrained by the export of dissipation from the equatorial Pacific.}
This dissipated-energy interpretation remains hypothetical, so a more realistic model is needed to test its validity and to further explore the physical connection between energy and entropy.

\FloatBarrier

\subsection{Future work} \label{subsec:future-work}

A natural first extension of the SRO toward a more realistic ENSO model is to incorporate nonlinear terms, state-dependent (multiplicative) noise, or red noise \citetext{\citealp{Levine+Jin2010JAS,Levine+2016,Kim+An2020GRL}; \citetalias{Kim+An2021JCLI}; \citealp{Han+2026JCLI}}.
TURs have also been formulated in broader stochastic settings that may be relevant to such extensions \citep{Otsubo+2020PRE,Terlizzi+Baiesi2020JPA,Tanogami+2023PRR,Han+2025PNAS}.
Applying these results to the SRO, however, would require reformulating the variance of the current and the partial entropy production rate.
Moreover, ENSO may be a noisy self-sustained oscillator rather than a damped one \citep{Weeks+Tziperman2025GRL}.
In that case, a TUR-like inequality could still constrain the coherence of the resulting stochastic limit cycle through its entropy production \citep{Santolin+Falasco2025PRL}.

Another extension is to multivariable systems that include other ocean basins or atmospheric variables.
For example, linear inverse models or extended ROs can capture the influence of interbasin interactions on ENSO \citep{Zhao+2024Nature,Lien+2026JAS}.
Full-system TURs have been obtained for multivariable systems \citep{Dechant2019JPA,Otsubo+2020PRE}, but the subsystem TUR used in our study was derived only for the bivariate case \citep{Tanogami+2023PRR}, so further extension would be needed.

Finally, TURs hold for arbitrary high-dimensional systems \citep{Dechant2019JPA,Otsubo+2020PRE} and may apply to gridded ensemble data from general circulation models.
This assumes that the resolved variables can be modeled as a Markov process.
A practical challenge is that the entropy production rate requires the probability density function, which is difficult to estimate from finite ensemble data in high dimensions.
The TUR can instead be used to estimate the entropy production rate from the mean and variance of a suitably chosen current \citep{Otsubo+2020PRE}.
These extensions would broaden the applicability of entropy-based constraints from idealized ROs to the full complexity of climate variability.

\section{Conclusions} \label{sec:conclusions}

\revA{We analyzed the SSTA variance $\langle T^2 \rangle$ in the linear SRO by using the TUR, which is a general inequality that constrains fluctuations even in nonlinear systems \citep[e.g.,][]{Horowitz+Gingrich2020NatPhys,Tanogami+2023PRR}.
Since the TUR is an inequality (i.e., not an equation), it does not determine the time evolution of $\langle T^2 \rangle$ but gives an upper bound on its tendency in terms of the partial entropy production rate $\dot{\cal P}$.
A sufficiently large $\dot{\cal P}$ is necessary for the seasonal change in $\langle T^2 \rangle$ that reflects phase locking.}
The dominant component of $\dot{\cal P}$ is the irreversibility indicator $\dot{\cal Q}$, given by the ratio of forward and backward transition probabilities of SSTA.
\revA{The seasonal variation of this irreversibility is governed by the growth rate $R(t)$, which reaches its maximum and minimum in boreal autumn and late winter, respectively.}
At both times, $\dot{\cal P}$ \revA{becomes large,} relaxing the TUR constraint on the SSTA variance.
\revA{The variance can then increase and decrease in boreal autumn and late winter, respectively, corresponding to the observed seasonal concentration of SSTA peaks.}
\revA{We also showed that $\dot{\cal P}$ splits into the irreversibility associated mainly with the seasonal change in $R(t)$ and that associated with the ENSO cycle, namely the recharge and discharge of equatorial heat content.
These results provide an entropy-based perspective on phase locking that complements the energy-based one \citepalias[e.g.,][]{Kim+An2021JCLI}.}

We further \revA{discussed} a \revA{hypothesis} for real ENSO.
If $\dot{\cal Q}$ were assumed to correspond to dissipated energy, the constraint on ENSO growth and decay would require this dissipation to be exported from the equatorial Pacific.
Unlike standard settings in nonequilibrium physics, however, it is unclear whether the noise source of \revA{ENSO} can be interpreted as a heat bath, so this interpretation remains hypothetical.
A more realistic model is needed to test this hypothesis and to further explore the physical connection between entropy and dissipated energy.
Beyond the SRO, the fluctuations intrinsic to ENSO \revA{could} be studied from the fundamental standpoint of nonequilibrium physics, whose principles apply to many systems rather than to ENSO alone.
This perspective may open a route toward a deeper understanding of variability across the climate system.

\clearpage
\acknowledgments

Y. Yasuda was supported by the Japan Society for the Promotion of Science (JSPS) KAKENHI (JP26K07197).
T. Kohyama was supported by JSPS KAKENHI (JP23H01241, JP23K13169), the MEXT Program for the Advanced Studies of Climate Change Projection (SENTAN) (JPMXD0722680395), and the Sumitomo Foundation (Environmental Research Project).
The authors declare no conflicts of interest.

%
%
\datastatement

The data and source code that support the findings of this study are preserved at the Zenodo repository (\url{https://doi.org/10.5281/zenodo.24702390}) and developed openly at the GitHub repository (\url{https://github.com/YukiYasuda2718/enso_sro_stochastic_constraint}).


\appendix[A] 
\appendixtitle{Re-derivation of the TUR} \label{app-sec:tur}

We re-derive the instantaneous TUR of \citet{Tanogami+2023PRR} using the short-time-limit method of \citet{Otsubo+2020PRE}.
We first set up the problem and then introduce the second law of information thermodynamics \citep{Horowitz+Esposito14PhysRevX,Loos+Klapp20NewJPhys} (Appendix~A\ref{app-subsec:second-law}).
By applying the Cauchy--Schwarz inequality to this second law, we obtain the TUR based on a general weight function $g(T,h)$ (Appendix~A\ref{app-subsec:tur-general}).
We then derive the TUR for the case $g(T,h) = T$ (Appendix~A\ref{app-subsec:tur-specific}).
We finally show that $\dot{\cal Q}$ is given by the ratio of the forward and backward transition probabilities (Appendix~A\ref{app-subsec:q}).

\subsection{Second law of information thermodynamics} \label{app-subsec:second-law}

We consider the following bivariate stochastic differential equation (SDE):
\begin{align}
    {\rm d} T &= -a^T(T, h, t) {\rm d} t + \sigma^T {\rm d} W^T, \label{eq:sde-t}\\
    {\rm d} h &= -a^h(T, h, t) {\rm d} t + \sigma^h {\rm d} W^h. \label{eq:sde-h} 
\end{align}
As in the SRO, $T$ and $h$ denote the sea surface temperature anomaly and thermocline depth anomaly, respectively.
$a^T$ and $a^h$ are nonlinear functions that may explicitly depend on $t$, and $W^T$ and $W^h$ are mutually independent standard Wiener processes.
They are related to the Gaussian white noises $\xi^T$ and $\xi^h$ by ${\rm d} W^T = \xi^T {\rm d}t$ and ${\rm d} W^h = \xi^h {\rm d}t$.
This system includes the linear SRO [Eqs.~(\ref{eq:linear-sro1})--(\ref{eq:linear-sro5})] as a special case.
For example, setting $-a^T = R(t) T + F_1 h$ recovers Eq.~(\ref{eq:linear-sro1}).
The TUR holds independently of the specific form of the drift, and thus we consider a general nonlinear SDE here.
We assume additive noise (that is, $\sigma^T$ and $\sigma^h$ are constants), as in the SRO analyzed in this study.
TURs can also be formulated in broader Langevin settings \citep{Otsubo+2020PRE,Tanogami+2023PRR}, but multiplicative noise would require re-deriving the probability flux, current variance, and explicit partial-entropy-production terms.

For this SDE, the joint probability density function $p^{T,h}$ of the state variables $(T,h)$ obeys the Fokker--Planck equation \citep[e.g.,][]{Gardiner09Book}:
\begin{align}
    \frac{\partial p^{T,h}}{\partial t} &= -\frac{\partial J^T}{\partial T} -\frac{\partial J^h}{\partial h}, \\
    p^{T,h} &:= p(T,h,t), \\
    J^T &:= -a^T p^{T,h}-\frac{\left(\sigma^T\right)^2}{2} \frac{\partial p^{T,h}}{\partial {T}},\label{eq:def-flux-jt} \\
    J^h &:= -a^h p^{T,h}-\frac{\left(\sigma^h\right)^2}{2} \frac{\partial p^{T,h}}{\partial {h}}.\label{eq:def-flux-jh}
\end{align}
This is a continuity equation in the $(T,h)$ state space.
The time derivative of $p^{T,h}$ equals minus the divergence of the probability flux $(J^T, J^h)$.
In the main text, we write $p^{T,h}$ as $p_t(T_t, h_t)$ [e.g., Eq.~(\ref{eq:def-m})].
When the coefficients of the SDE explicitly depend on $t$ (for example, $R(t)$ in the SRO), $p^{T,h}$ can also explicitly depend on $t$.
The subscript $t$ in $p_t$ expresses this time dependence.
We impose the boundary condition that the probability flux vanishes at infinity in the $(T,h)$ space.
In what follows, we derive the TUR for $T$; analogous formulas hold for $h$.

The following inequality, the \emph{second law of information thermodynamics}, holds for $T$ \citep{Horowitz+Esposito14PhysRevX,Loos+Klapp20NewJPhys}.
For a detailed review, see \citet{Yasuda+Kohyama2025JCLI}.
Although that review assumes that $a^T$ and $a^h$ do not explicitly depend on $t$, this assumption is not used in the derivation, so the same argument extends to time-dependent $a^T$ and $a^h$:
\begin{align}
    \dot{{\cal P}} := \frac{2\dot{\cal Q}}{\left(\sigma^T\right)^2}+\frac{{\rm d} {\cal S}}{{\rm d} t}-\dot{{\cal I}}&=\left\langle\frac{2J^T}{p^{T,h} \left(\sigma^T\right)^2} \circ \frac{{\rm d} T}{{\rm d} t}\right\rangle=\int {\rm d}T{\rm d}h \; \frac{2\left(J^T\right)^2}{p^{T,h} \left(\sigma^T\right)^2} \geq 0. \label{eq:second-law-in-app}
\end{align}
$\dot{{\cal P}}$ is called the partial entropy production rate and is always nonnegative \citep{Allahverdyan+09JStatMech,Shiraishi+Sagawa2015,Rosinberg+Horowitz2016}.
Equality holds when $J^T = 0$, that is, when the probability flux associated with the $T$ transitions vanishes \citep{Loos+Klapp20NewJPhys}.
The quantities in the inequality are defined as follows:
\begin{align}
    \dot{\cal Q}&:=\left\langle -a^T \circ \frac{{\rm d} T}{{\rm d} t}\right\rangle, \label{eq:def-q-in-app} \\
    {\cal S} &:=-\int {\rm d} T \; p(T,t) \ln p(T,t), \label{eq:def-S-in-app} \\
    p(T,t) &:=\int {\rm d}h \; p(T, h, t), \label{eq:def-marginal-p-T}\\
    \dot{{\cal I}} &:= \int {\rm d}T{\rm d}h \; \frac{\partial J^T}{\partial T} \ln \left[ \frac{p(T,t)}{p(T,h,t)}\right]. \label{eq:def-m2}
\end{align}
$\dot{\cal Q}$ is an indicator of irreversibility \citep{Kurchan1998JPA,Seifert2005PRL,Sekimoto10Book,Ito16Book} and is further discussed in Appendix~A\ref{app-subsec:q}.
The Shannon entropy ${\cal S}$ represents the uncertainty of the continuous variable $T$, that is, a differential entropy \citep[e.g.,][]{Cover+Thomas05Book}.
In the main text, we write $p(T,t)$ as $p_t(T_t)$ [e.g., Eq.~(\ref{eq:def-s})].
$\dot{{\cal I}}$ is called the information flow from $h$ to $T$ and represents the contribution to the rate of change of the correlation between $T$ and $h$ induced by the time evolution of $T$ \citep{Allahverdyan+09JStatMech,Horowitz+Esposito14PhysRevX,Rosinberg+Horowitz2016,Loos+Klapp20NewJPhys}.
$\dot{{\cal I}}$ is not the total time derivative of the mutual information, because that total tendency also includes the contribution induced by the time evolution of $h$.
Equation~(\ref{eq:def-i}) may seem more intuitive, but Eq.~(\ref{eq:def-m2}) is more general because it also holds for multivariable systems \citep{Loos+Klapp20NewJPhys}.

\subsection{TUR based on a general weight function $g(T,h)$} \label{app-subsec:tur-general}

We derive the TUR for the following current ${\cal J}$ based on a general weight function:
\begin{align}
    {\cal J} &:= g(T,h) \circ {\rm d}T \label{eq:def-current-general}\\
    &= \left[ -a^T g + \frac{\left(\sigma^T\right)^2}{2} \frac{\partial g}{\partial T} \right] {\rm d} t + g \sigma^T {\rm d} W^T, \label{eq:def-current-general-expanded}
\end{align}
where we substitute the definition of the Stratonovich product, perform a Taylor expansion, and neglect terms of higher order in ${\rm d}t$.

\revA{To simplify the derivation, we introduce the \emph{mean local velocity} of $T$ \citep{Chetrite+Gawedzki2009JSP, Seifert12RepProgPhys}:
\begin{align}
    u^T := \frac{J^T}{p^{T,h}} = -a^T - \frac{\left(\sigma^T\right)^2}{2} \frac{\partial \ln p^{T,h}}{\partial T}, \label{eq:def-mean-local-velocity}
\end{align}
where the second equality follows from Eq.~(\ref{eq:def-flux-jt}).
Equivalently, $J^T = p^{T,h} u^T$, which has the same form as a mass flux (density times flow velocity) in fluid dynamics.
$u^T$ is the velocity at which the probability density is transported along $T$ in the state space of $(T,h)$ \citep{Chetrite+Gawedzki2009JSP}.
Note that $u^T$ is not the velocity of a fluid and $u^T$ has the same units as ${\rm d}T/{\rm d}t$.
As a preliminary, we use $u^T$ to write $\dot{\cal P}$ in Eq.~(\ref{eq:second-law-in-app}) in a simpler form:
\begin{align}
    \dot{\cal P} = \int {\rm d}T{\rm d}h \; \frac{2\left(J^T\right)^2}{p^{T,h} \left(\sigma^T\right)^2} = \frac{2}{\left(\sigma^T\right)^2} \left\langle \left(u^T\right)^2 \right\rangle. \label{eq:pdot-in-app}
\end{align}

}%
The mean of the current then becomes \citep{Otsubo+2020PRE}:
\revA{\begin{align}
    \left \langle {\cal J} \right \rangle &= {\rm d} t \left \langle -a^T g + \frac{\left(\sigma^T\right)^2}{2} \frac{\partial g}{\partial T} \right \rangle \\
    &= {\rm d} t \int {\rm d}T{\rm d}h \; \left[ -a^T g + \frac{\left(\sigma^T\right)^2}{2} \frac{\partial g}{\partial T} \right] p^{T,h} \\
    &= {\rm d} t \int {\rm d}T{\rm d}h \; gJ^T \\
    &= {\rm d} t \left\langle g u^T \right\rangle. \label{eq:mean-current-general-in-app}
\end{align}}%
The third equality follows by integration by parts, expressing the result in terms of the probability flux $J^T$ in Eq.~(\ref{eq:def-flux-jt}), \revA{and the last follows from Eq.~(\ref{eq:def-mean-local-velocity}).}
Next, the variance of the current becomes \citep{Otsubo+2020PRE}:
\revA{\begin{align}
    {\rm Var}[{\cal J}] &:= \left \langle {\cal J}^2 \right \rangle - \left \langle {\cal J} \right \rangle^2 \\
    &= \left \langle {\cal J}^2 \right \rangle + O\left(\left({\rm d}t\right)^2\right) \\
    &= {\rm d} t \left[\int {\rm d}T{\rm d}h \; \left(\sigma^T\right)^2 g^2 p^{T,h}\right] \\
    &= \left(\sigma^T\right)^2 \left\langle g^2 \right\rangle {\rm d}t. \label{eq:variance-current-general-in-app}
\end{align}}%
Here, we use $({\rm d} W^T)^2 = {\rm d}t$ \citep[e.g.,][]{Gardiner09Book} and neglect terms of second order or higher in ${\rm d}t$.

With these preparations, the TUR follows by applying the Cauchy--Schwarz inequality \citep{Otsubo+2020PRE}:
\revA{\begin{align}
    {\rm Var}[{\cal J}] \dot{\cal P} \; {\rm d} t &= 2 ({\rm d} t)^2 \left\langle g^2 \right\rangle \left\langle \left(u^T\right)^2 \right\rangle \\
    &= 2({\rm d} t)^2 \left[ \int {\rm d}T{\rm d}h \; \left(g \sqrt{p^{T,h}}\right)^2 \right] \left[ \int {\rm d}T{\rm d}h \; \left(u^T \sqrt{p^{T,h}}\right)^2 \right] \\
    &\ge 2 ({\rm d} t)^2 \left[ \int {\rm d}T{\rm d}h \; \left(g \sqrt{p^{T,h}}\right) \; \left(u^T \sqrt{p^{T,h}}\right) \right]^2 \\
    &= 2 ({\rm d} t)^2 \left\langle g u^T \right\rangle^2 \\
    &= 2\left \langle {\cal J} \right \rangle^2,
\end{align}}%
\revA{where we use Eqs.~(\ref{eq:pdot-in-app}), (\ref{eq:mean-current-general-in-app}), and~(\ref{eq:variance-current-general-in-app}).
Here, functions of $(T,h)$ are regarded as vectors whose inner product is the ensemble average $\langle \cdot \rangle$ of their product.
Equality holds if and only if the two functions $g$ and $u^T$ are \emph{parallel}; that is, $u^T(T,h,t) = c(t)\,g(T,h)$ for all $(T,h)$ with some $c(t)$.}
Assuming ${\rm Var}[{\cal J}] > 0$, we obtain the following TUR \citep{Tanogami+2023PRR}:
\begin{align}
    \frac{\dot{\cal P}}{2} \ge  \frac{\left \langle {\cal J} \right \rangle^2}{{\rm Var}[{\cal J}] \; {\rm d} t}. \label{eq:tur-in-appendix}
\end{align}
The right-hand side is the squared mean of the current normalized by the variance, corresponding to a signal-to-noise ratio.
If this ratio is regarded as a measure of precision, then the TUR can be interpreted as stating that increasing precision requires paying the thermodynamic cost of increasing the partial entropy production rate $\dot{\cal P}$.
For the full system $(T,h)$, a TUR can be derived in the same way \citep{Otsubo+2020PRE}, but a characteristic feature of the subsystem TUR for $T$, Eq.~(\ref{eq:tur-in-appendix}), is that $\dot{\cal P}$ contains the information flow $\dot{\cal I}$.
This means that information transfer from $h$ to $T$ can modify the upper bound on the precision \citep{Tanogami+2023PRR}.

\subsection{TUR based on the specific weight function $g(T,h)=T$} \label{app-subsec:tur-specific}

We derive the TUR for the case in which the weight function is $g(T,h)=T$.
First, the mean and variance of the current ${\cal J}$ become:
\begin{align}
    \left \langle {\cal J} \right \rangle &= \left\langle T \circ {\rm d} T \right\rangle = \frac{1}{2} \left\langle {\rm d}(T^2) \right\rangle = \frac{1}{2} \left\langle \frac{{\rm d} T^2}{{\rm d}t} \right\rangle {\rm d}t = \frac{1}{2} \left( \frac{{\rm d}}{{\rm d}t} \left\langle T^2 \right\rangle \right) {\rm d}t, \label{eq:mean-current-specific-in-app} \\
    {\rm Var}[{\cal J}] &= \left[\int {\rm d}T{\rm d}h \; \left(\sigma^T\right)^2 T^2 p^{T,h}\right] {\rm d}t = \left(\sigma^T\right)^2\left\langle T^2 \right\rangle {\rm d}t.\label{eq:variance-current-specific-in-app}
\end{align}
In Eq.~(\ref{eq:mean-current-specific-in-app}), we use the fact that the Stratonovich product satisfies the chain rule of differentiation \citep[e.g.,][]{Gardiner09Book}.
In general, the current ${\cal J}$ represents a net displacement, but by using $T$ as the weight function, its mean $\left \langle {\cal J} \right \rangle$ becomes the net change in $\left\langle T^2 \right\rangle$.
Positive and negative changes are both possible, and the sign corresponds to the direction of the current.
In Eq.~(\ref{eq:variance-current-specific-in-app}), the variance of the current, ${\rm Var}[{\cal J}]$, involves the squared noise amplitude $\left(\sigma^T\right)^2$ and the mean square of $T$.

Thus, the TUR becomes the following inequality:
\begin{align}
    \frac{\dot{\cal P}}{2} \ge \frac{1}{\left(\sigma^T\right)^2\left\langle T^2 \right\rangle} \left( \frac{1}{2} \frac{{\rm d}}{{\rm d}t}  \left\langle T^2 \right\rangle \right)^2. \label{eq:tur-in-appendix2}
\end{align}
$\dot{\cal P}$ provides an upper bound on the normalized squared tendency of $\left\langle T^2 \right\rangle$.
After this normalization, a large tendency of $\left\langle T^2 \right\rangle$ must be accompanied by a sufficiently large $\dot{\cal P}$.
In this derivation, we neither use the specific form of the SRO's governing equations nor assume any specific initial condition.
The derived TUR holds at any time, for any initial condition, and for any $T$ obeying an arbitrary nonlinear SDE with additive noise, provided that the current variance is nonzero.
For the specific current used here, this condition requires $\sigma^T>0$ and $\left\langle T^2 \right\rangle>0$.

\subsection{$\dot{\cal Q}$ as a ratio of transition probabilities} \label{app-subsec:q}

We show that $\dot{\cal Q}$ is given by a ratio of transition probabilities \citep{Sekimoto10Book,Seifert12RepProgPhys,Ito16Book}.
First, the probability density function of ${\rm d} W^T$ is a Gaussian \citep[e.g.,][]{Shiraishi23Book}:
\begin{equation}
    p({\rm d} W^T) = \frac{1}{\sqrt{2\pi {\rm d}t}} \exp\left\{ -\frac{1}{2} \frac{\left({\rm d} W^T\right)^2}{{\rm d}t} \right\}.
\end{equation}
This expression holds for an infinitesimal increment ${\rm d} W^T$ at any time.
Suppose that the state at time $t$ is $(T_t, h_t)$.
By discretizing the governing equation~(\ref{eq:sde-t}) for $T$ using the Euler--Maruyama method and solving for ${\rm d} W^T$, we obtain the transition density of $T$ from $p({\rm d} W^T)$.
The noise coefficient matrix in the governing equations [Eqs.~(\ref{eq:sde-t}) and~(\ref{eq:sde-h})] is diagonal, and thus the noise ${\rm d} W^T$ contributes only to the transition of $T$ \citep[the bipartite condition; e.g.,][]{Horowitz+Esposito14PhysRevX}:
\begin{align}
    p_t(T_{t+{\rm d}t} \mid T_t, h_t) = \frac{1}{\sqrt{2\pi \left(\sigma^T\right)^2 {\rm d}t}} \exp\left\{-\frac{1}{2} \frac{\left[T_{t+{\rm d}t} - T_t + a^T(T_t, h_t, t){\rm d}t\right]^2}{\left(\sigma^T\right)^2{\rm d}t}\right\}.
\end{align}
Here, we set ${\rm d} T = T_{t+{\rm d}t} - T_t$ and use the fact that the Jacobian of the variable transformation is $1/\sigma^T$.
When $a^T$ explicitly depends on $t$, $p_t$ also explicitly depends on $t$.
We next consider the backward transition.
The variable $T$ does not change sign under time reversal, and thus the backward transition density is given by
\begin{align}
    p_t(T_t \mid T_{t+{\rm d}t}, h_t) = \frac{1}{\sqrt{2\pi \left(\sigma^T\right)^2 {\rm d}t}} \exp\left\{-\frac{1}{2} \frac{\left[T_t - T_{t+{\rm d}t} + a^T(T_{t+{\rm d}t}, h_t, t){\rm d}t\right]^2}{\left(\sigma^T\right)^2{\rm d}t}\right\}.
\end{align}

Taking the logarithm of the ratio of the forward and backward transition densities yields:
\begin{align}
    \ln \frac{p_t(T_{t+{\rm d}t} \mid T_t, h_t)}{p_t(T_t \mid T_{t+{\rm d}t}, h_t)} &= -\frac{2}{\left(\sigma^T\right)^2}\frac{a^T(T_t, h_t, t) + a^T(T_{t+{\rm d}t}, h_t, t)}{2} \left( T_{t+{\rm d}t} - T_t \right) + O({\rm d}t) \\
    &= -\frac{2}{\left(\sigma^T\right)^2}\frac{a^T(T_t, h_t, t) + a^T(T_{t+{\rm d}t}, h_{t+{\rm d}t}, t+{\rm d}t)}{2} \left( T_{t+{\rm d}t} - T_t \right) + O({\rm d}t) \\
    &= - \frac{2}{\left(\sigma^T\right)^2} a^T(T_t, h_t, t) \circ {\rm d}T + O({\rm d}t).
\end{align}
Taking the ensemble average and using the definition of $\dot{\cal Q}$ [Eq.~(\ref{eq:def-q-in-app})], we obtain \citep{Sekimoto10Book,Seifert12RepProgPhys,Ito16Book}:
\begin{align}
    \left\langle \ln \frac{p_t(T_{t+{\rm d}t} \mid T_t, h_t)}{p_t(T_t \mid T_{t+{\rm d}t}, h_t)} \right\rangle &= \frac{2}{\left(\sigma^T\right)^2} \dot{\cal Q} {\rm d}t.
\end{align}
The neglected $O({\rm d}t)$ term becomes $o({\rm d}t)$ through the ensemble average and thus does not contribute to $\dot{\cal Q}$.
The quantity $\dot{\cal Q}$ is proportional to the logarithm of the ratio of the forward and backward transition probabilities.
The more likely the forward transition is relative to the backward transition (that is, the more irreversible the transition), the larger $\dot{\cal Q}$.
When the noise source is a heat bath in equilibrium, $\dot{\cal Q}$ equals the heat per unit time dissipated from the system to the bath \citep{Sekimoto10Book,Seifert12RepProgPhys,Ito16Book}.
However, when the noise source is not a heat bath (i.e., noise is not thermal), it is not clear whether $\dot{\cal Q}$ can be interpreted as heat.
In Section~\ref{sec:discussion}\ref{subsec:interpretation}, we discuss the interpretation of $\dot{\cal Q}$ for real ENSO.

\appendix[B] 
\appendixtitle{Expressions for stochastic-thermodynamic quantities of the SRO} \label{app-sec:expression-sro}

This appendix presents the expressions for the stochastic-thermodynamic quantities based on the SRO's governing equations [Eqs.~(\ref{eq:linear-sro1})--(\ref{eq:linear-sro5})].
For general bivariate linear SDEs with time-dependent coefficients, these quantities were derived by \citet{Ito+Sagawa15NatComm} and \citet{Matsumoto+Sagawa2018PRE}.
Here, we present only those used in this study.

The stochastic-thermodynamic quantities (e.g., $\dot{\cal Q}$) are given by the means and covariance matrix of $T$ and $h$.
We therefore discuss the time evolution of these statistical quantities.
The evolution equations for the means are obtained by taking the ensemble average of the SRO [Eqs.~(\ref{eq:linear-sro1}) and~(\ref{eq:linear-sro2})]:
\begin{align}
    \frac{{\rm d}\left\langle T \right\rangle}{{\rm d} t}&=R(t) \left\langle T \right\rangle +F_1 \left\langle h \right\rangle, \\
    \frac{{\rm d}\left\langle h \right\rangle}{{\rm d} t}&=-\varepsilon \left\langle h\right\rangle -F_2 \left\langle T \right\rangle.
\end{align}
The means at each time $t$ are obtained by numerically solving these equations.
Since the SRO is a damped oscillator (Section~\ref{sec:sro}), the means relax to zero.
Next, we denote the covariance matrix by $\mathsf{\Sigma}$:
\begin{equation}
    \mathsf{\Sigma} :=\left(
      \begin{array}{cc}
        \Sigma^{TT} & \Sigma^{Th} \\
        \Sigma^{hT} & \Sigma^{hh}
      \end{array}
    \right),
\end{equation}
where $\Sigma^{Th}=\Sigma^{hT}$.
Using the linearity of the SRO, the time evolution of $\mathsf{\Sigma}$ is given by \citep[e.g.,][]{Gardiner09Book,Matsumoto+Sagawa2018PRE}:
\begin{align}
    \frac{{\rm d} \mathsf{\Sigma}}{{\rm d} t} &= -\mathsf{A}\mathsf{\Sigma} - \mathsf{\Sigma}\mathsf{A}^{\rm T} + \mathsf{B}, \label{eq:covariance-evolution} \\
    \mathsf{A} &:= \left(
      \begin{array}{cc}
        -R(t) & -F_1 \\
        F_2 & \varepsilon
      \end{array}
    \right), \\
    \mathsf{B} &:= \left(
      \begin{array}{cc}
        \left(\sigma^{T}\right)^2 & 0 \\
        0 & \left(\sigma^h\right)^2
      \end{array}
    \right).
\end{align}
The covariance matrix at each time $t$ is obtained by numerically solving these equations.

For Figs.~\ref{fig:tur-KA21}, \ref{fig:decomposition-q-KA21}, and~\ref{fig:decomposition-q-H26}, we integrated the mean and covariance equations with the explicit Euler method using a time step of $1/360$~yr for 200~yr.
The initial mean was $\left(\langle T\rangle,\langle h\rangle\right)=(0~{\rm K},10~{\rm m})$.
The initial covariance matrix was the steady covariance matrix obtained by fixing the coefficients at the initial time.
We then used the last 360 time steps as the final model year.

After spin-up, this covariance matrix determines the covariance-only expressions for the stochastic-thermodynamic quantities used in the TUR.
This reduction assumes that $(T,h)$ is in the cyclostationary Gaussian state reached by the linear additive-noise SRO after the effect of the initial condition has decayed \citep[e.g.,][]{Gardiner09Book}.
\revA{Here, we show the resulting expressions for $\dot{\cal Q}$, ${\rm d}{\cal S}/{\rm d}t$, and $\dot{{\cal I}}$ \citep{Matsumoto+Sagawa2018PRE}.}
\begin{align}
    \dot{\cal Q} &= -\left\langle a^T \circ \frac{{\rm d} T}{{\rm d} t}\right\rangle \\
    &= \left\langle (a^T)^2 \right\rangle - \left\langle \frac{\partial a^T}{\partial T} \right\rangle \frac{\left(\sigma^T\right)^2}{2} \\
    &= R(t)^2 \left\langle T^2 \right\rangle + 2 R(t) F_1 \left\langle T h \right\rangle + \left(F_1\right)^2 \left\langle h^2 \right\rangle + \frac{R(t)\left(\sigma^T\right)^2}{2}.
\end{align}
\revA{This derivation starts from the definition of $\dot{\cal Q}$ in Eq.~(\ref{eq:def-q-in-app}).
For the second equality, we perform a Taylor expansion of the Stratonovich product, as in Eq.~(\ref{eq:def-current-general-expanded}).
The third equality follows from $-a^T = R(t) T + F_1 h$.}

\revA{The Shannon entropy in Eq.~(\ref{eq:def-S-in-app}) becomes ${\cal S} = (1/2) \ln\left(2\pi\Sigma^{TT}\right) + 1/2$ for the Gaussian density $p(T,t)$ [$=p^T$] with variance $\Sigma^{TT}$.
The time derivative of ${\cal S}$ is}
\begin{align}
    \frac{{\rm d} {\cal S}}{{\rm d} t} &= \frac{1}{2\Sigma^{TT}} \frac{{\rm d} \Sigma^{TT}}{{\rm d} t} \label{eq:dSdt-1} \\
    &= R(t) + \frac{F_1\Sigma^{Th}}{\Sigma^{TT}} + \frac{\left(\sigma^T\right)^2}{2\Sigma^{TT}}, \label{eq:dSdt-2}
\end{align}
\revA{where the second equality follows from the $TT$ component of Eq.~(\ref{eq:covariance-evolution}).}

\revA{We obtain the information flow $\dot{\cal I}$ by integrating Eq.~(\ref{eq:def-m2}) by parts and substituting Eq.~(\ref{eq:def-flux-jt}) and the Gaussian densities $p^{T}$ and $p^{T,h}$ with zero mean:
\begin{align}
    \dot{{\cal I}} &= -\int {\rm d}T{\rm d}h \; J^T \frac{\partial}{\partial T} \left(\ln \frac{p^{T}}{p^{T,h}}\right)\\
    &= -\int {\rm d}T{\rm d}h \; \left\{ \left[ R(t) T + F_1 h \right] p^{T,h} - \frac{\left(\sigma^T\right)^2}{2} \frac{\partial p^{T,h}}{\partial T} \right\} \left( \frac{\Sigma^{hh} T - \Sigma^{Th} h}{\det \mathsf{\Sigma}} - \frac{T}{\Sigma^{TT}} \right) \\
    &= -\frac{\Sigma^{hh}}{\det \mathsf{\Sigma}} \left[ R(t) \Sigma^{TT} + F_1 \Sigma^{Th} \right] + \frac{\Sigma^{Th}}{\det \mathsf{\Sigma}} \left[ R(t) \Sigma^{Th} + F_1 \Sigma^{hh} \right] - \frac{\left(\sigma^T\right)^2 \Sigma^{hh}}{2\det \mathsf{\Sigma}} \nonumber \\
    &\quad + \frac{1}{\Sigma^{TT}} \left[ R(t) \Sigma^{TT} + F_1 \Sigma^{Th} \right] + \frac{\left(\sigma^T\right)^2}{2\Sigma^{TT}} \\
    &= \frac{F_1\Sigma^{Th}}{\Sigma^{TT}} - \frac{\left(\sigma^T\right)^2 \Sigma^{hh}}{2\det \mathsf{\Sigma}} + \frac{\left(\sigma^T\right)^2}{2\Sigma^{TT}},
\end{align}
where the third equality uses integration by parts for the term with $\partial p^{T,h}/\partial T$, and for the last equality, the terms with $R(t)$ cancel.}

In the above expressions, the explicit time dependence is written only for the growth rate $R(t)$, but the means and covariance matrix also depend on $t$.
These quantities ($\dot{\cal Q}$, ${\rm d}{\cal S}/{\rm d}t$, and $\dot{{\cal I}}$) yield the partial entropy production rate $\dot{\cal P}$ via Eq.~(\ref{eq:second-law-in-app}), and the TUR~(\ref{eq:tur-in-appendix2}) can then be evaluated.

\appendix[C] 
\appendixtitle{\revA{Equality in the TUR and the split of $\dot{\cal P}$}} \label{app-sec:tur-equality}

\subsection{\revA{TUR in the one-variable model}} \label{app-subsec:one-variable}

\revA{We consider the one-variable model in Section~\ref{sec:tur}\ref{subsec:tur-one-variable}, obtained from the SRO [Eqs.~(\ref{eq:linear-sro1})--(\ref{eq:linear-sro5})] by setting $F_1 = 0$ and discarding $h$:
\begin{align}
    \frac{{\rm d} T}{{\rm d} t} = R(t) T + \sigma^T \xi^T. \label{eq:one-variable-model}
\end{align}
The results of Appendix~A hold for this model, provided that $p^{T,h}$ [$= p(T,h,t)$] is replaced by $p^T$ [$=p(T,t)$].
Note that $\dot{\cal I}$ in Eq.~(\ref{eq:def-m2}) is zero because $h$ is absent.
We assume that $R_0 < 0$ and that the influence of the initial condition has sufficiently decayed.
Then, by the linearity of Eq.~(\ref{eq:one-variable-model}), $T$ follows a Gaussian distribution with zero mean and variance $\left\langle T^2 \right\rangle$ \citep[e.g.,][]{Gardiner09Book}, which gives $\partial \ln p/\partial T = -T/\left\langle T^2 \right\rangle$.
Substituting this and $-a^T = R(t) T$ into Eq.~(\ref{eq:def-mean-local-velocity}), we obtain
\begin{align}
    u^T = \left[ R(t) + \frac{\left(\sigma^T\right)^2}{2\left\langle T^2 \right\rangle} \right] T = \frac{{\rm d} {\cal S}}{{\rm d} t} T, \label{eq:u-one-variable}
\end{align}
where we have used Eq.~(\ref{eq:dSdt-2}) with $F_1 = 0$ and $\Sigma^{TT} = \left\langle T^2 \right\rangle$.
The coefficient (i.e., ${{\rm d} {\cal S}}/{{\rm d} t}$) is a function of $t$ alone because it is obtained through the ensemble average over $T$.

The mean local velocity $u^T$ is thus parallel to the weight function $g = T$, and the TUR~(\ref{eq:tur-in-appendix2}) holds with equality at every $t$.
Indeed, Eqs.~(\ref{eq:pdot-in-app}) and~(\ref{eq:u-one-variable}) yield
\begin{align}
    \frac{\dot{\cal P}}{2} = \frac{\left\langle T^2 \right\rangle}{\left(\sigma^T\right)^2} \left( \frac{{\rm d} {\cal S}}{{\rm d} t} \right)^2 = \frac{1}{\left(\sigma^T\right)^2 \left\langle T^2 \right\rangle} \left( \frac{1}{2}\frac{{\rm d}}{{\rm d}t} \left\langle T^2 \right\rangle \right)^2, \label{eq:tur-equality-one-variable}
\end{align}
where we have used Eq.~(\ref{eq:dSdt-1}) with $\Sigma^{TT} = \left\langle T^2 \right\rangle$.

We consider two cases, classified according to $R_a$.
In the first case, when $R_a = 0$, the growth rate is the negative constant $R_0$.
The variance also becomes constant, $\left\langle T^2 \right\rangle = -\left(\sigma^T\right)^2/(2R_0)$, where the damping by $R_0$ balances the fluctuations driven by the noise.
Then ${\rm d}{\cal S}/{\rm d}t = 0$, and hence $u^T = 0$ at all $T$, leading to $\dot{\cal P} = 0$.
In the other case, when $R_a \neq 0$, the growth rate varies seasonally.
The variance $\left\langle T^2 \right\rangle$ also varies seasonally, and $\dot{\cal P} > 0$ except at the extrema of $\left\langle T^2 \right\rangle$, where ${\rm d}{\cal S}/{\rm d}t = 0$.

We next show that $\dot{\cal P}$ is given by the ratio of the probabilities of the forward and backward (i.e., time-reversed) processes, whereas $\dot{\cal Q}$ is given by the ratio of the transition probabilities alone (Section~\ref{sec:tur}\ref{subsec:tur-one-variable}).
Over a short interval from $t$ to $t+{\rm d}t$, the probability of the forward process is the product of the probability density $p_t(T_t)$ of the initial state and the transition density $p_t(T_{t+{\rm d}t} \mid T_t)$.
The time-reversed process starts from $T_{t+{\rm d}t}$, whose probability density is $p_{t+{\rm d}t}(T_{t+{\rm d}t})$, and returns to $T_t$ with the backward transition density $p_t(T_t \mid T_{t+{\rm d}t})$.
The ensemble average of the logarithm of the ratio of these two probabilities is
\begin{align}
    \left\langle \ln \frac{p_t(T_t)\, p_t(T_{t+{\rm d}t} \mid T_t)}{p_{t+{\rm d}t}(T_{t+{\rm d}t})\, p_t(T_t \mid T_{t+{\rm d}t})} \right\rangle
    &= \left\langle \ln \frac{p_t(T_{t+{\rm d}t} \mid T_t)}{p_t(T_t \mid T_{t+{\rm d}t})} \right\rangle + \left\langle \ln \frac{p_t(T_t)}{p_{t+{\rm d}t}(T_{t+{\rm d}t})} \right\rangle \nonumber \\
    &= \frac{2\dot{\cal Q}}{\left(\sigma^T\right)^2} {\rm d}t + \frac{{\rm d}{\cal S}}{{\rm d}t} {\rm d}t \nonumber \\
    &= \dot{\cal P} \, {\rm d}t, \label{eq:pdot-process-ratio}
\end{align}
where the first term follows from Appendix~A\ref{app-subsec:q}, the second from ${\cal S} = -\left\langle \ln p_t(T_t) \right\rangle$ in Eq.~(\ref{eq:def-S-in-app}) evaluated at $t$ and $t+{\rm d}t$, and the last equality from Eq.~(\ref{eq:second-law-in-app}) with $\dot{\cal I} = 0$.

\subsection{\revA{TUR in the two-variable model}} \label{app-subsec:two-variable}

We next consider the SRO with both $T$ and $h$ [Eqs.~(\ref{eq:linear-sro1})--(\ref{eq:linear-sro5})] and reveal the deviation of the TUR~(\ref{eq:tur-in-appendix2}) from equality.
First, we regress the mean local velocity $u^T$ on $T$:
\begin{align}
    u^T &= c_\parallel(t)\,T + r, \label{eq:regression-u}\\
    c_\parallel(t) &:= \frac{\left\langle T u^T \right\rangle}{\left\langle T^2 \right\rangle}, \label{eq:regression-coeff}
\end{align}
where the residual $r$ is uncorrelated with $T$ by construction, $\left\langle T r \right\rangle = 0$.
As in the one-variable model [see Eq.~(\ref{eq:u-one-variable})], $c_\parallel$ is given by
\begin{align}
    c_\parallel(t) = \frac{1}{2\left\langle T^2 \right\rangle}\frac{{\rm d}\left\langle T^2 \right\rangle}{{\rm d}t} = \frac{{\rm d}{\cal S}}{{\rm d}t}, \label{eq:c-parallel-by-entropy}
\end{align}
where we have used Eqs.~(\ref{eq:mean-current-general-in-app}), (\ref{eq:mean-current-specific-in-app}), and~(\ref{eq:dSdt-1}) with $g = T$ and $\Sigma^{TT} = \left\langle T^2 \right\rangle$.
Substituting Eq.~(\ref{eq:regression-u}) into Eq.~(\ref{eq:pdot-in-app}) and using $\left\langle T r \right\rangle = 0$, we obtain
\begin{align}
    \dot{\cal P} = \frac{2}{\left(\sigma^T\right)^2 \left\langle T^2 \right\rangle} \left\langle T \circ \frac{{\rm d}T}{{\rm d}t} \right\rangle^2 + \frac{2}{\left(\sigma^T\right)^2} \left\langle r^2 \right\rangle. \label{eq:pdot-regression}
\end{align}
The first term equals the right-hand side of the TUR~(\ref{eq:tur-main2}) [or twice Eq.~(\ref{eq:tur-in-appendix2})] because $\left\langle T \circ {\rm d}T \right\rangle = (1/2)\,({\rm d}\left\langle T^2 \right\rangle/{\rm d}t)\,{\rm d}t$ in Eq.~(\ref{eq:mean-current-specific-in-app}).
The second term is the deviation of the TUR~(\ref{eq:tur-main2}) from equality.
It is nonnegative and vanishes only when $r = 0$, that is, when $u^T$ is parallel to $T$, as in the one-variable model (Appendix~C\ref{app-subsec:one-variable}).

We now derive the explicit form of $r$.
In the SRO, $u^T$ in Eq.~(\ref{eq:def-mean-local-velocity}) is a linear combination of $T$ and $h$.
This is because $-a^T = R(t) T + F_1 h$ and $p^{T,h}$ is Gaussian once the influence of the initial condition has decayed (Appendix~B).
Furthermore, $h$ is decomposed into the part correlated with $T$ and the uncorrelated part $h_\perp$:
\begin{align}
    h_\perp := h - \frac{\left\langle T h \right\rangle}{\left\langle T^2 \right\rangle} T. \label{eq:def-h-perp}
\end{align}
By definition, $\left\langle T h_\perp \right\rangle = 0$.
Then, $u^T$ can be written as a linear combination of $T$ and $h_\perp$, and hence the residual $r$ in Eq.~(\ref{eq:regression-u}) is proportional to $h_\perp$:
\begin{align}
    r &= c_\perp(t)\,h_\perp, \label{eq:regression-r}\\
    c_\perp(t) &:= \frac{\left\langle h_\perp u^T \right\rangle}{\left\langle h_\perp^2 \right\rangle}. \label{eq:regression-coeff2}
\end{align}

Using this form of $r$, we finally decompose $\dot{\cal P}$ in Eq.~(\ref{eq:pdot-regression}) as follows:
\begin{align}
    \dot{\cal P} &= \dot{\cal P}_{\rm var} + \dot{\cal P}_{\rm cyc}, \label{eq:split-in-app} \\
    \dot{\cal P}_{\rm var} &:= \frac{2}{\left(\sigma^T\right)^2\left\langle T^2 \right\rangle} \left\langle T \circ \frac{{\rm d}T}{{\rm d}t} \right\rangle^2, \label{eq:def-pdot-var} \\
    \dot{\cal P}_{\rm cyc} &:= \frac{2}{\left(\sigma^T\right)^2\left\langle h_\perp^2 \right\rangle} \left\langle h_\perp \circ \frac{{\rm d}T}{{\rm d}t} \right\rangle^2, \label{eq:def-pdot-osc}
\end{align}
where we have used Eqs.~(\ref{eq:def-current-general}) and~(\ref{eq:mean-current-general-in-app}) with $g = h_\perp$, together with $\left\langle r^2 \right\rangle = \left\langle h_\perp u^T \right\rangle^2/\left\langle h_\perp^2 \right\rangle$ from Eqs.~(\ref{eq:regression-r}) and~(\ref{eq:regression-coeff2}).

For $R_a = 0$, we have $\dot{\cal P}_{\rm var} = 0$ and $\dot{\cal P} = \dot{\cal P}_{\rm cyc}$ because the SRO reaches a steady state, in which $\left\langle T^2 \right\rangle$ is constant.
$\dot{\cal P}_{\rm cyc}$ is positive because the coupling terms $F_1 h$ and $-F_2 T$ in Eqs.~(\ref{eq:linear-sro1}) and~(\ref{eq:linear-sro2}) have opposite signs (i.e., $F_1 > 0$ and $-F_2 < 0$), which breaks detailed balance \citep{Weiss07,Loos+Klapp20NewJPhys}.
Therefore, unlike in the one-variable model, equality in the TUR does not hold even without seasonality.

\subsection{\revA{Explicit forms of $c_\parallel$ and $c_\perp$}} \label{app-subsec:pdot-osc-explicit}

To examine the seasonality of $\dot{\cal P}_{\rm var}$ and $\dot{\cal P}_{\rm cyc}$, we derive the explicit forms of $c_\parallel$ and $c_\perp$ in terms of the SRO coefficients such as $R(t)$ and $F_1$.
First, by combining Eqs.~(\ref{eq:regression-u}) and~(\ref{eq:regression-r}), the mean local velocity $u^T$ can be written with $c_\parallel$ and $c_\perp$:
\begin{align}
    u^T = c_\parallel(t)\,T + c_\perp(t)\,h_\perp. \label{eq:u-regression-sro}
\end{align}
Recall that $u^T$ is defined in Eq.~(\ref{eq:def-mean-local-velocity}).
Since $(T,h)$ follows a Gaussian distribution with zero mean and covariance matrix $\mathsf{\Sigma}$ (Appendix~B), we have
\begin{align}
    \frac{\partial \ln p^{T,h}}{\partial T} = -\frac{\Sigma^{hh} T - \Sigma^{Th} h}{\det \mathsf{\Sigma}}, \label{eq:score-func-T}
\end{align}
where we have used the explicit form of $\mathsf{\Sigma}^{-1}$.
Substituting $-a^T = R(t) T + F_1 h$ and Eq.~(\ref{eq:score-func-T}) into Eq.~(\ref{eq:def-mean-local-velocity}), we obtain $u^T$ as a linear combination of $T$ and $h$:
\begin{align}
    u^T = \left[ R(t) + \frac{\left(\sigma^T\right)^2 \Sigma^{hh}}{2\det \mathsf{\Sigma}} \right] T + \left[ F_1 - \frac{\left(\sigma^T\right)^2 \Sigma^{Th}}{2\det \mathsf{\Sigma}} \right] h. \label{eq:u-linear-t-h}
\end{align}
Here, $h = \left(\Sigma^{Th}/\Sigma^{TT}\right) T + h_\perp$ holds by Eq.~(\ref{eq:def-h-perp}), $\left\langle T^2 \right\rangle = \Sigma^{TT}$, and $\left\langle T h \right\rangle = \Sigma^{Th}$.
Comparing Eq.~(\ref{eq:u-linear-t-h}) with Eq.~(\ref{eq:u-regression-sro}), we obtain
\begin{align}
    c_\parallel(t) &= R(t) + \frac{F_1 \Sigma^{Th}}{\Sigma^{TT}} + \frac{\left(\sigma^T\right)^2}{2\Sigma^{TT}}, \label{eq:c-parallel-explicit} \\
    c_\perp(t) &= F_1 - \frac{\left(\sigma^T\right)^2 \Sigma^{Th}}{2\det \mathsf{\Sigma}}, \label{eq:c-perp-explicit}
\end{align}
where we have used $\det \mathsf{\Sigma} = \Sigma^{TT}\Sigma^{hh} - \left(\Sigma^{Th}\right)^2$.
Equation~(\ref{eq:c-parallel-explicit}) is consistent with Eq.~(\ref{eq:c-parallel-by-entropy}) via Eq.~(\ref{eq:dSdt-2}).

Finally, we express $\dot{\cal P}_{\rm var}$ and $\dot{\cal P}_{\rm cyc}$ in terms of $c_\parallel$ and $c_\perp$.
Using Eqs.~(\ref{eq:pdot-in-app}) and~(\ref{eq:u-regression-sro}), together with $\left\langle T h_\perp \right\rangle = 0$, we obtain
\begin{align}
    \dot{\cal P}_{\rm var} &= \frac{2}{\left(\sigma^T\right)^2} c_\parallel^2 \left\langle T^2 \right\rangle, \label{eq:pdot-var-explicit} \\
    \dot{\cal P}_{\rm cyc} &= \frac{2}{\left(\sigma^T\right)^2} c_\perp^2 \left\langle h_\perp^2 \right\rangle, \label{eq:pdot-cyc-explicit}
\end{align}
which are equivalent to Eqs.~(\ref{eq:def-pdot-var}) and~(\ref{eq:def-pdot-osc}), respectively.
These expressions reveal that the growth rate $R(t)$ appears explicitly in $\dot{\cal P}_{\rm var}$ through $c_\parallel$ but not in $\dot{\cal P}_{\rm cyc}$.
Thus, $\dot{\cal P}_{\rm cyc}$ varies seasonally only implicitly through the covariance matrix $\mathsf{\Sigma}$.

Equation~(\ref{eq:pdot-cyc-explicit}) also relates $\dot{\cal P}_{\rm cyc}$ to the term $C$ in Eq.~(\ref{eq:decompose-q-a-to-d}).
Using the correlation coefficient $\rho := \Sigma^{Th}/\sqrt{\Sigma^{TT}\Sigma^{hh}}$, we have $\left\langle h_\perp^2 \right\rangle = \left(1-\rho^2\right)\Sigma^{hh}$, and hence
\begin{align}
    \dot{\cal P}_{\rm cyc} = \left(\frac{c_\perp}{F_1}\right)^2 \left(1-\rho^2\right) C. \label{eq:pdot-cyc-term-c}
\end{align}
When $T$ and $h$ are uncorrelated ($\rho = 0$), $c_\perp = F_1$ holds [Eq.~(\ref{eq:c-perp-explicit})] and $\dot{\cal P}_{\rm cyc}$ equals $C$ itself.
Importantly, the associations of $\dot{\cal P}_{\rm cyc}$ with the ENSO cycle and of $\dot{\cal P}_{\rm var}$ with the delayed seasonal response hold exactly only in two limits: the one-variable model, in which $\dot{\cal P}_{\rm cyc} = 0$ (Appendix~C\ref{app-subsec:one-variable}), and the SRO with $R_a = 0$, in which $\dot{\cal P}_{\rm var} = 0$ (Appendix~C\ref{app-subsec:two-variable}).
In general, $\dot{\cal P}_{\rm var}$ includes the contribution of the ENSO cycle through $F_1\Sigma^{Th}/\Sigma^{TT}$ in $c_\parallel$ [Eq.~(\ref{eq:c-parallel-explicit})], whereas $\dot{\cal P}_{\rm cyc}$ is also affected by the seasonality [i.e., $R(t)$] implicitly through $\mathsf{\Sigma}$.}

\appendix[D] 
\appendixtitle{Results for $h$ in the western equatorial Pacific} \label{app-sec:results-H26}

We discuss the results obtained with the parameter set of \citet{Han+2026JCLI}, in which $h$ is taken as the thermocline depth anomaly in the western equatorial Pacific (Table~\ref{table:sro-parameters}).
In the main text, following \citetalias{Kim+An2021JCLI}, $h$ is taken as the thermocline depth anomaly over the entire equatorial Pacific.
Here, we show only the figures that differ qualitatively from those in the main text.

\begin{figure}[t]
    \centering
    \noindent\includegraphics[width=14cm]{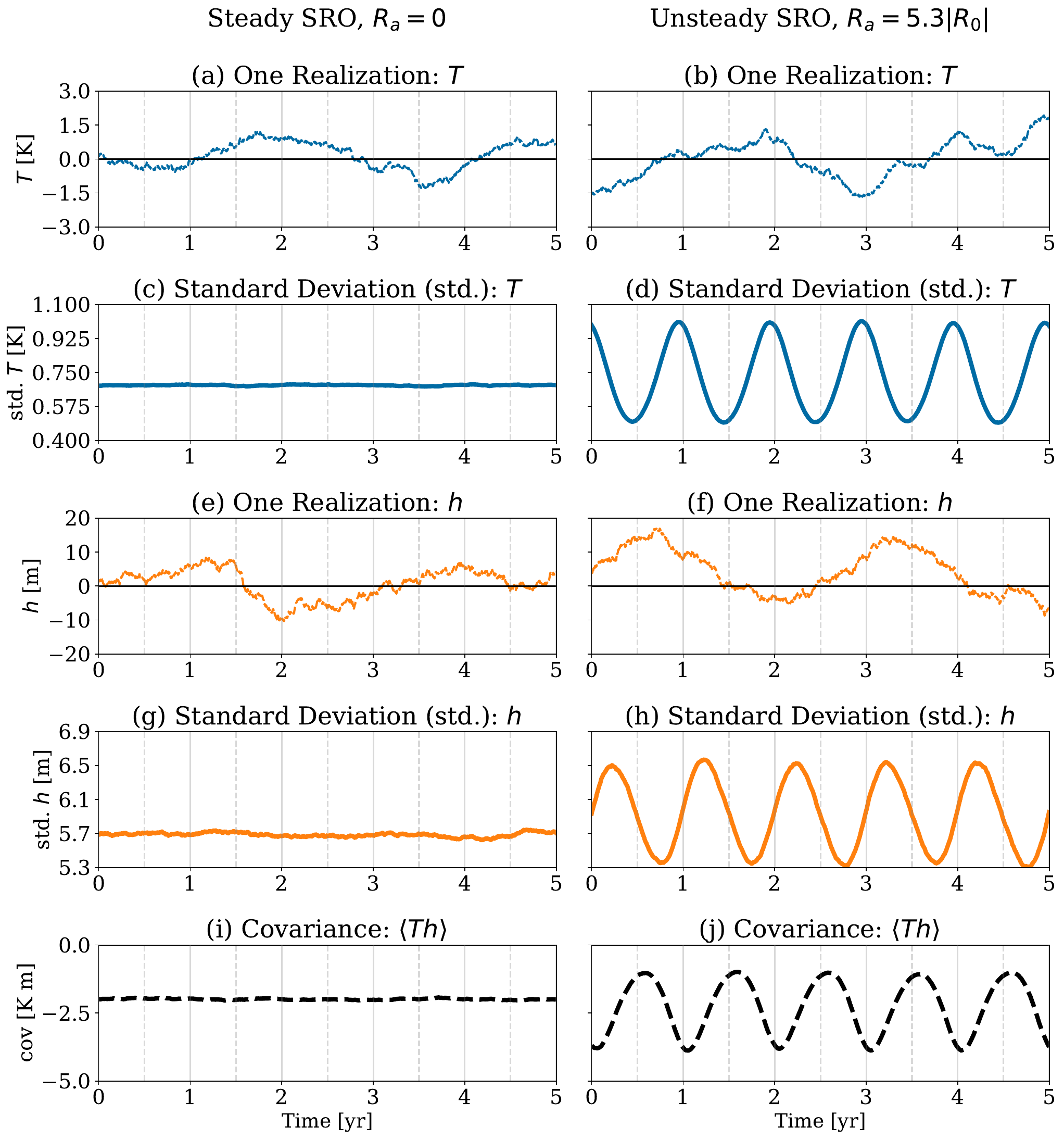}
    \caption{
        Simulation results of the SRO with the parameter set of \citet{Han+2026JCLI}, where $h$ is the thermocline depth anomaly in the western equatorial Pacific (Table~\ref{table:sro-parameters}).
        \revA{The left column shows the steady SRO with $R_a = 0$, whereas the right shows the unsteady SRO with $R_a = 5.3 \lvert R_0 \rvert$.}
        All other settings are the same as in Fig.~\ref{fig:sro-time-series-KA21}.
    }
    \label{fig:sro-time-series-H26}
\end{figure}

Figure~\ref{fig:sro-time-series-H26} shows the simulation results of the SRO.
The main difference from Fig.~\ref{fig:sro-time-series-KA21} is the sign of the covariance $\left\langle T h \right\rangle$, which is always negative regardless of the time dependence of $R$ (Fig.~\ref{fig:sro-time-series-H26}i,j).
This negative covariance reflects a statistical tendency of the trajectory in the state space of $(T,h)$ \citep{Han+2026JCLI}.
The trajectory $(T_t, h_t)$ tends to have a negative slope when $h$ is taken as the thermocline depth anomaly in the western equatorial Pacific, but a positive slope when taken over the entire equatorial Pacific.
See \citet{Han+2026JCLI} for details.

\FloatBarrier

\begin{figure}[t]
    \centering
    \noindent\includegraphics[width=16.45cm]{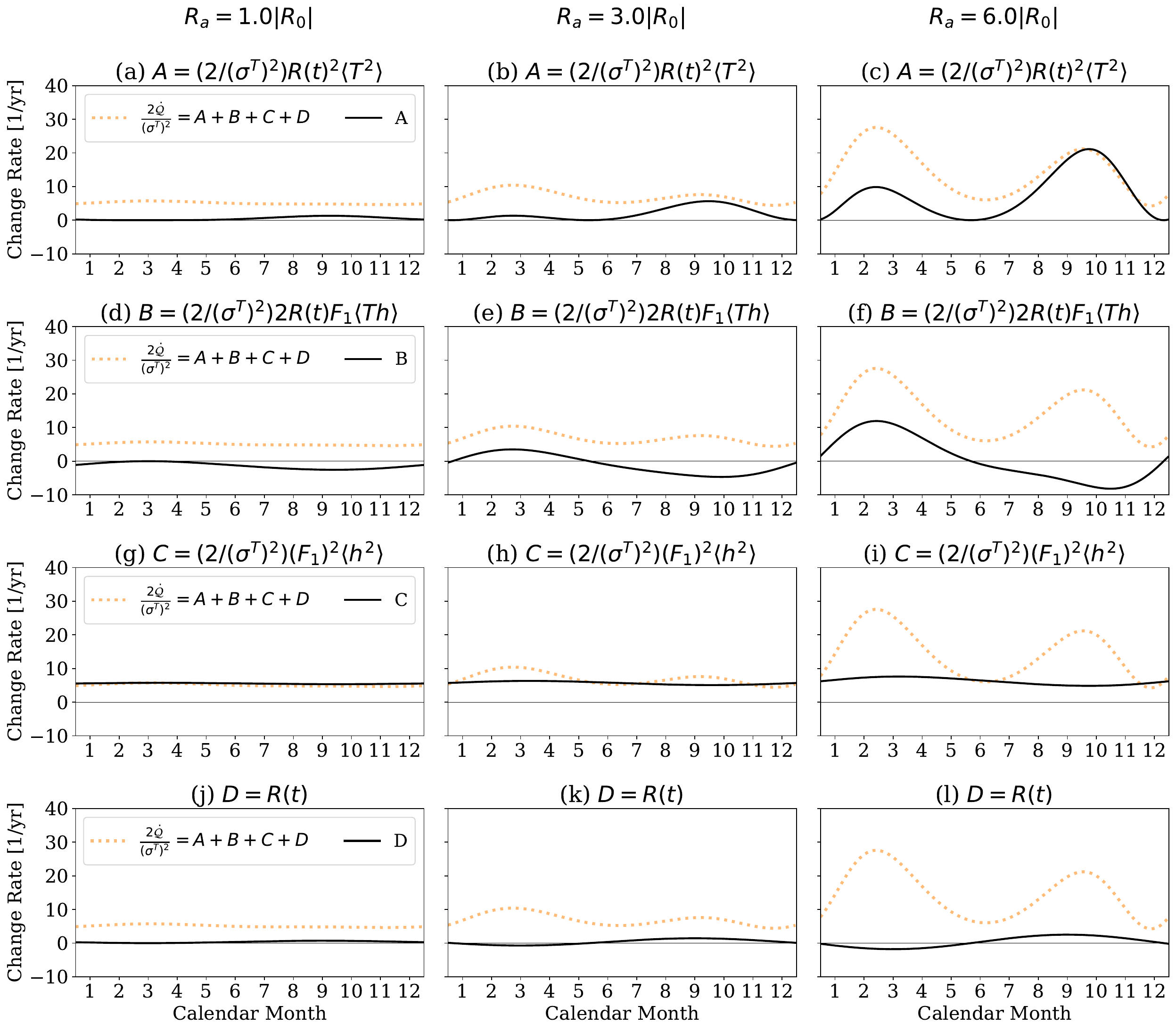}
    \caption{
        Time series of the terms $A$, $B$, $C$, and $D$ of the irreversibility indicator ${2\dot{\cal Q}}/{\left(\sigma^T\right)^2}$ in Eq.~(\ref{eq:decompose-q-a-to-d}) with the parameter set of \citet{Han+2026JCLI}, where $h$ is the thermocline depth anomaly in the western equatorial Pacific (Table~\ref{table:sro-parameters}).
        The largest amplitude, $R_a = 6.0 \lvert R_0 \rvert$, is close to the actual H26 ratio $R_a \simeq 5.3 \lvert R_0 \rvert$ used in Fig.~\ref{fig:sro-time-series-H26}.
        All other settings are the same as in Fig.~\ref{fig:decomposition-q-KA21}.
    }
    \label{fig:decomposition-q-H26}
\end{figure}

Figure~\ref{fig:decomposition-q-H26} shows the time series of $A$--$D$ composing ${2\dot{\cal Q}}/{\left(\sigma^T\right)^2}$ in Eq.~(\ref{eq:decompose-q-a-to-d}).
As in Fig.~\ref{fig:decomposition-q-KA21}, ${2\dot{\cal Q}}/{\left(\sigma^T\right)^2}$ exhibits a wavenumber-2 structure.
We again focus on $A$--$C$.
First, $C$ involves the thermocline-depth variance $\left\langle h^2 \right\rangle$, but its seasonality is weak (Fig.~\ref{fig:decomposition-q-H26}g--i).
This result suggests that the \revA{seasonal variation} of $\left\langle h^2 \right\rangle$ does not necessarily correspond to \revA{that} of irreversibility.

As in Fig.~\ref{fig:decomposition-q-KA21}, the seasonality of ${2\dot{\cal Q}}/{\left(\sigma^T\right)^2}$ is produced mainly by $A$ and $B$, but their relative contributions differ.
The term $A$ has a wavenumber-2 structure originating from $R(t)^2$ [Eq.~(\ref{eq:decompose-q-a-to-d}) and Fig.~\ref{fig:decomposition-q-H26}a--c].
The growth rate $R(t)$ attains its extrema around September and March, so $A$ peaks in both seasons, although the precise timing and amplitude are modulated by the variance $\langle T^2 \rangle$.
By contrast, the term $B$ has a wavenumber-1 structure (Fig.~\ref{fig:decomposition-q-H26}d--f), unlike its wavenumber-2 structure in Fig.~\ref{fig:decomposition-q-KA21}d--f.
The covariance $\left\langle T h \right\rangle$ is always negative; thus, $B$ in Eq.~(\ref{eq:decompose-q-a-to-d}) varies as $-R(t)$.
The physical picture of ENSO should not depend on the choice of variables.
Therefore, the difference between Figs.~\ref{fig:decomposition-q-KA21} and~\ref{fig:decomposition-q-H26} suggests that $A$ and $B$ should not be interpreted separately.
The combined contribution $A+B$ represents the seasonal variation of irreversibility associated with air--sea interaction in the SRO.
\revA{This contribution corresponds mainly to $\dot{\cal P}_{\rm var}$ in Eq.~(\ref{eq:split-main}) (Section~\ref{sec:results}).}

\FloatBarrier

\appendix[E] 
\appendixtitle{\revA{Dependence of $\Delta \dot{\cal P}$ on other SRO parameters}} \label{app-sec:fixed-ra}

\revA{We examine the dependence of $\Delta \dot{\cal P}$ (Section~\ref{sec:results}) on parameters other than $R_a$.
For the \citetalias{Kim+An2021JCLI} parameter set in Table~\ref{table:sro-parameters}, we fixed $R_a$ and varied $R_0$, $\varepsilon$, and $F_2$ separately: $R_0$ was shifted by $-2.4$ to $+0.8$~yr$^{-1}$, whereas $\varepsilon$ and $F_2$ were multiplied by factors from 0.0625 to 9.5 and from 0.25 to 4, respectively.
These parameters modify the ENSO period, which is estimated from the imaginary part of the BWJ index \citep{Jin+2006GRL,Lu+2018JCLI,Jin+2020inbook}.
Each case was integrated for 50,000 years, and El Ni\~no peaks were detected as in Fig.~\ref{fig:peak-seasonality-KA21}.
The peak months depend on the parameters, so we introduce the following metric to evaluate the concentration:
\begin{align}
    \bar r := \left\lvert \frac{1}{N} \sum_{j=1}^{N} e^{i\theta_j} \right\rvert. \label{eq:mean-resultant-length}
\end{align}
Here, $i$ is the imaginary unit, $N$ is the number of peaks, and $\theta_j$ is the calendar date of the $j$-th peak expressed as an angle on the annual cycle.
$\bar r$ is 0 for peaks distributed uniformly and 1 for peaks on a single date.

Figure~\ref{fig:fixed-ra-concentration} shows the dependence of $\bar r$ on $\Delta \dot{\cal P}$ and the ENSO period for each parameter sweep, including the $R_a$ sweep in Fig.~\ref{fig:peak-seasonality-KA21}.
Here, \emph{Base} denotes the result for the original \citetalias{Kim+An2021JCLI} parameters (Table~\ref{table:sro-parameters}).
Even when $R_a$ is fixed, $\bar r$ varies with the other parameters.
Up to the base value of $\Delta \dot{\cal P}$, $\bar r$ increases almost monotonically with $\Delta \dot{\cal P}$ and the values of $\bar r$ are close to the curve of the $R_a$ sweep (Fig.~\ref{fig:fixed-ra-concentration}a).
Above the base value, $\bar r$ still tends to increase with $\Delta \dot{\cal P}$, but the relation has branches that depend on the varied parameter.
Similar results were obtained with the parameter sets of \citet{Vialard+2025RevGeophys} and \citet{Han+2026JCLI} (not shown).

If $\Delta \dot{\cal P}$ is regarded as a phase-locking index, it has a different dependence on the parameters from that of the seasonal energy index (SEI; \citetalias{Kim+An2021JCLI}).
In our notation, the SEI is $\int [R(t) - R_0]\,{\rm d}t = (R_a/\omega)\cos(\omega t + \phi)$, which depends only on $R_a$ and takes the same value in all the above experiments, except for the $R_a$ sweep.
In contrast, $\Delta \dot{\cal P}$ reflects the other parameters such as $\varepsilon$, which is also clear from the theoretical discussion (Appendices~B and~C).
However, the relation between $\Delta \dot{\cal P}$ and $\bar r$ is not simple.
For example, when $F_2$ is varied, $\bar r$ has two branches (Fig.~\ref{fig:fixed-ra-concentration}a,b).
The branch for longer ENSO periods (smaller $F_2$) lies above the curve of the $R_a$ sweep, whereas that for shorter periods (larger $F_2$) lies below it.

\begin{figure}[t]
    \centering
    \noindent\includegraphics[width=14cm]{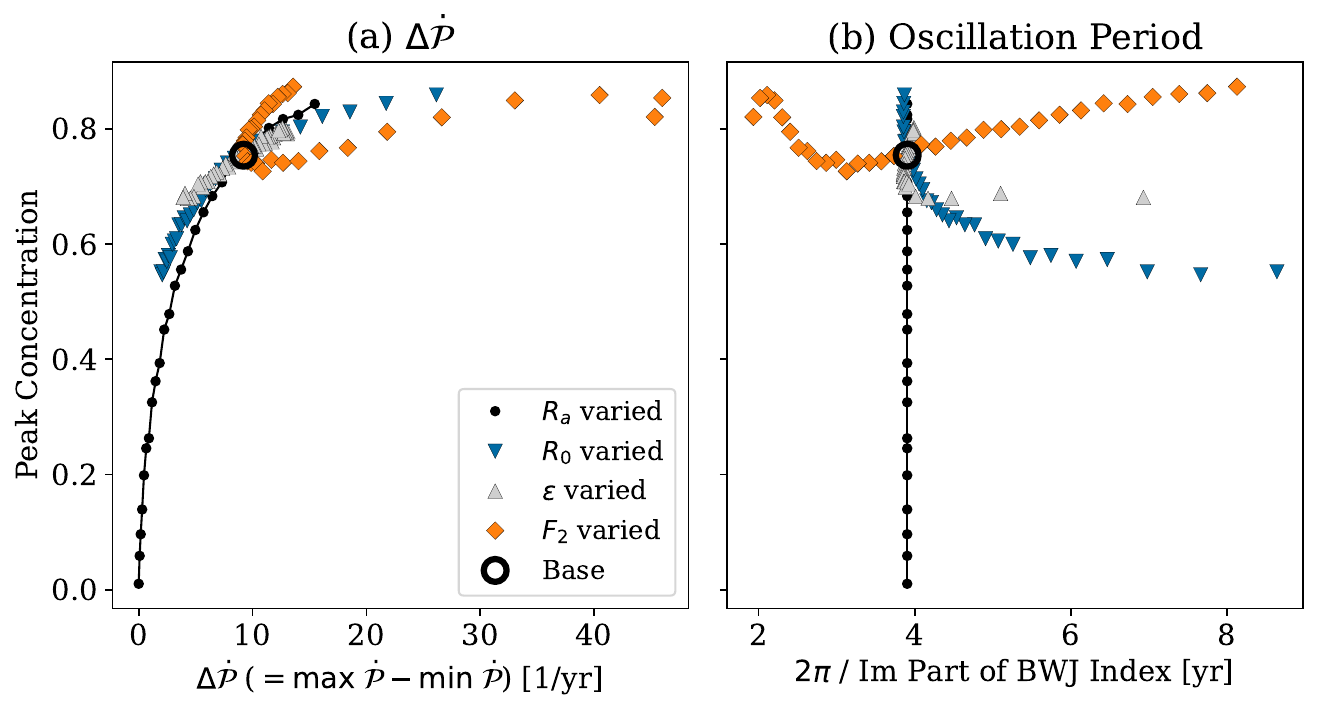}
    \caption{
        \revA{Concentration of El Ni\~no peaks in the SRO for the parameter set of \citetalias{Kim+An2021JCLI}.
        The concentration is measured by $\bar r$ in Eq.~(\ref{eq:mean-resultant-length}).
        In each sweep, we varied only one of $R_a$, $R_0$, $\varepsilon$, and $F_2$ (see the text for their ranges).
        Each case was integrated for 50,000 years, and El Ni\~no peaks were detected as in Fig.~\ref{fig:peak-seasonality-KA21}.
        (a) Concentration $\bar r$ against $\Delta \dot{\cal P}$.
        (b) Concentration $\bar r$ against the ENSO period, $2\pi$ divided by the imaginary part of the BWJ index.
        The label \emph{Base} denotes the result for the original \citetalias{Kim+An2021JCLI} parameters (Table~\ref{table:sro-parameters}).}
    }
    \label{fig:fixed-ra-concentration}
\end{figure}
\FloatBarrier

Figure~\ref{fig:fixed-ra-f2} shows the dependence on $F_2$ in a layout similar to that of Fig.~\ref{fig:peak-seasonality-KA21}.
As $F_2$ increases (i.e., the ENSO period shortens), $\Delta \dot{\cal P}$ rises sharply (Fig.~\ref{fig:fixed-ra-f2}a).
The peaks remain concentrated but shift slightly earlier, toward October--November (Fig.~\ref{fig:fixed-ra-f2}b), consistent with previous studies that discuss the dependence of phase locking on the ENSO period in the RO, including parametric resonance near a 2-yr period \citep{Stein+2014JCLI,Chen+Jin2020JCLI}.
With a shorter period, the ratio $\Delta \dot{\cal P}_{\rm var}/\Delta \dot{\cal P}$ decreases (Fig.~\ref{fig:fixed-ra-f2}c), where $\Delta \dot{\cal P}_{\rm var} := \max \;\dot{\cal P}_{\rm var} - \min \;\dot{\cal P}_{\rm var}$.
This decrease can be interpreted as a relative increase in $\dot{\cal P}_{\rm cyc}$, so that $\Delta \dot{\cal P}$ no longer mainly reflects $\dot{\cal P}_{\rm var}$.
Thus, the deviation from the curve of the $R_a$ sweep at shorter periods likely reflects the relative increase in $\dot{\cal P}_{\rm cyc}$.

\begin{figure}[t]
    \centering
    \noindent\includegraphics[width=16.45cm]{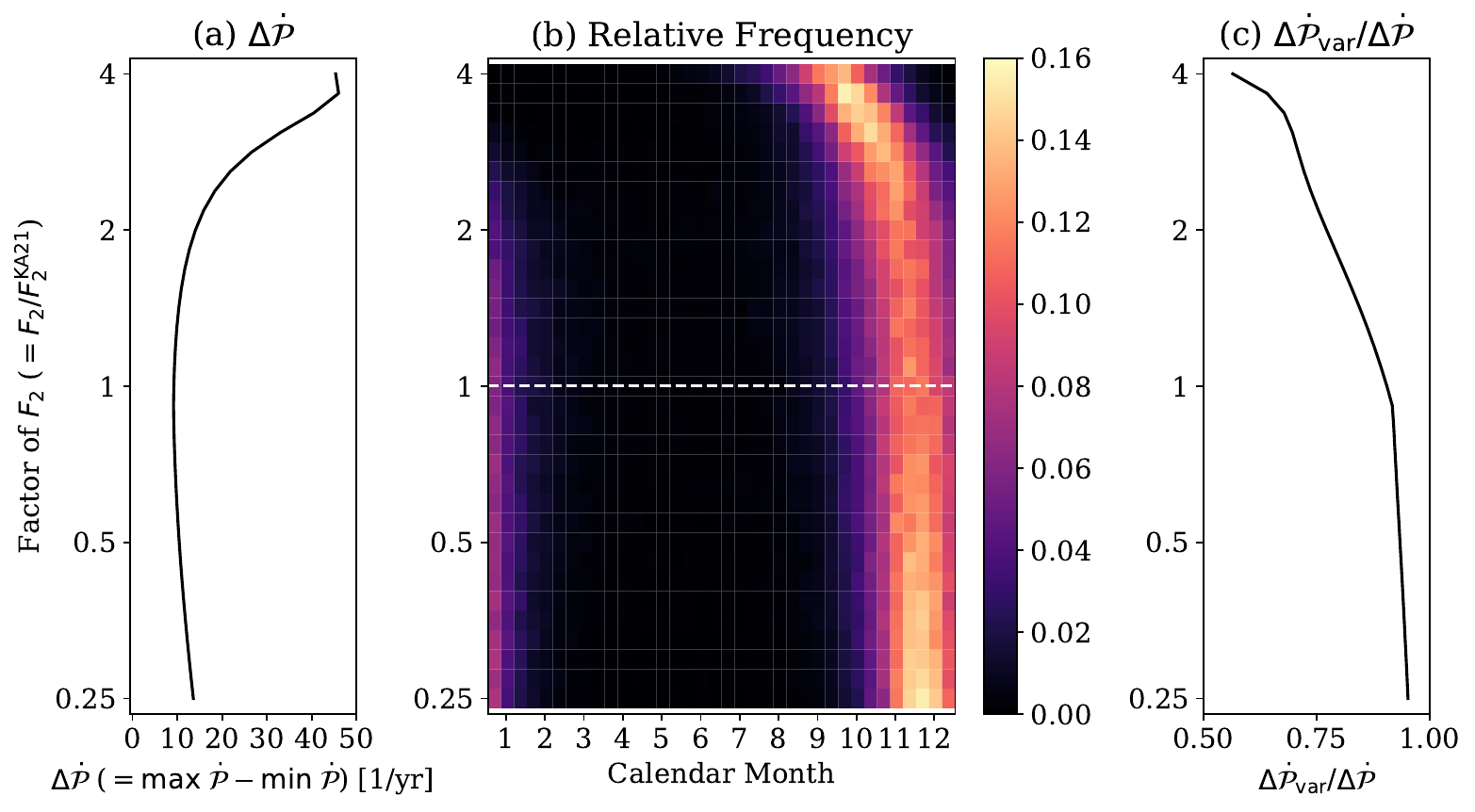}
    \caption{
        \revA{Distributions of El Ni\~no peaks in the SRO when $F_2$ is varied for the parameter set of \citetalias{Kim+An2021JCLI} in Table~\ref{table:sro-parameters} ($R_a = 2.0 \lvert R_0 \rvert$).
        The vertical axis is the factor multiplying $F_2$.
        (a) $\Delta \dot{\cal P}$.
        (b) Relative frequency of peaks across calendar months.
        (c) Ratio $\Delta \dot{\cal P}_{\rm var}/\Delta \dot{\cal P}$, where $\Delta \dot{\cal P}_{\rm var}$ is the range of $\dot{\cal P}_{\rm var}$.
        All other settings are the same as in Fig.~\ref{fig:peak-seasonality-KA21}, except for the color scale of (b).}
    }
    \label{fig:fixed-ra-f2}
\end{figure}

Even for large $F_2$ (e.g., a factor of 4), the TUR~(\ref{eq:tur-main2}) still constrains the tendency of $\left\langle T^2 \right\rangle$, and the time series of $\dot{\cal P}$ broadly follows that of the right-hand side, as in Fig.~\ref{fig:tur-KA21} (not shown).
The time series of $\dot{\cal P}$ itself, rather than a quantity extracted from it such as $\Delta \dot{\cal P}$, should therefore be examined first.}

\FloatBarrier

\bibliographystyle{ametsocV6}
\bibliography{references}

\end{document}